\documentclass[preprint,12pt]{elsarticle}
\usepackage{natbib}
\usepackage{epsf}
\usepackage{timet}
\usepackage{amsmath}
\usepackage{mathtools}
\usepackage{mdwlist}
\usepackage{setspace}
\usepackage{amsfonts}
\usepackage{amssymb}
\usepackage{graphicx}   
\usepackage{subfigure}
\usepackage{float}
\usepackage[english]{babel} %language selection
\usepackage{hyperref}
\hypersetup{colorlinks, 
           citecolor=black,
           filecolor=black,
           linkcolor=red,
           urlcolor=blue,
           bookmarksopen=true,
           }
\usepackage{enumerate}
\usepackage[dvipsnames]{xcolor}
\newcommand{\tp}{\mathrm{p}}
\newcommand{\ts}{\mathrm{s}}
\newcommand{\tsh}{\mathrm{sh}}
\newcommand{\tsv}{\mathrm{sv}}
\newcommand{\ud}{\,\mathrm{d}}
\newcommand{\FGA}{\mathrm{F}}
\newcommand{\nb}{\nabla}
\newcommand{\TR}{\mathrm{tr}}
\newcommand{\RE}{\mathrm{re}}
\newcommand{\f}[2]{\frac{#1}{#2}}
\newcommand{\IN}{\mathrm{in}}

\newcommand{\p}{\partial}

\newcommand{\abs}[1]{\lvert#1\rvert}

\newcommand{\bd}[1]{\boldsymbol{#1}}
\newcommand{\I}{\mathrm{i}}

\newcommand{\bF}{\mathbf F}

\newcommand{\mR}{\mathbb R}

\newcommand{\bff}{\mathbf f}

\newcommand{\bg}{\mathbf g}

\newcommand{\bN}{\mathbf{N}}

\newcommand{\bu}{\mathbf u}
\newcommand{\bp}{\mathbf{p}}
\newcommand{\bP}{\mathbf{P}}

\newcommand{\bQ}{\mathbf{Q}}
\newcommand{\bx}{\mathbf{x}}

\newcommand{\bz}{\mathbf{z}}

\DeclareMathOperator{\sgn}{sgn}

\newcommand{\x}{\ _\Box}
\newcommand{\ds}{\displaystyle}
\usepackage{pdfpages} % to include book cover in pdf format.
\reversemarginpar
\newcounter{mnote}
\begin{document}
%\title[Deep Learning Seismic Substructure Detection Using the FGA]{Deep Learning Seismic Substructure Detection Using the Frozen Gaussian Approximation}

\title{Deep Learning Seismic Substructure Detection Using the Frozen Gaussian Approximation}
 
%\author[J. C. Hateley, J. Roberts, K. Mylonakis, X. Yang]{James C. Hateley, Jay Roberts, Kyle Mylonakis, Xu Yang \\
%Department of Mathematics, University of California, Santa Barbara, CA 93106, USA}
%\date{}
%\pagerange{}
%\volume{}
%\pubyear{}
%\ead{hateleyjc@\allowbreak gmail.\allowbreak com, jayroberts@\allowbreak math.\allowbreak ucsb.\allowbreak edu, kmylonakis@\allowbreak math.\allowbreak ucsb.\allowbreak edu, xuyang@\allowbreak math.\allowbreak ucsb.\allowbreak edu}

\author{James C. Hateley}\ead{hateleyjc@gmail.com}
\author{Jay Roberts}\ead{jayroberts@math.ucsb.edu}
\author{ Kyle Mylonakis}\ead{ kmylonakis@math.ucsb.edu}
\author{Xu Yang}\ead{xuyang@math.ucsb.edu}

%\author[1]{James C. Hateley\fnref{fn1}}\ead{hateleyjc@gmail.com}
%\author[2]{Jay Roberts\fnref{fn2}}\ead{jayroberts@math.ucsb.edu}
%\author[3]{ Kyle Mylonakis\fnref{fn3}}\ead{ kmylonakis@math.ucsb.edu}
%\author[4]{Xu Yang\corref{cor1}\fnref{fn4}}\ead{xuyang@math.ucsb.edu}

\cortext[cor1]{Corresponding author}
%\fntext[fn1]{author footnote.}
%\fntext[fn2]{Another author footnote, this is a very longfootnote and it should be a really long footnote. But thisfootnote is not yet sufficiently long enough to make twolines of footnote text.}
%\fntext[fn3]{author footnote.}
%\fntext[fn4]{author footnote.}

\address{Department of Mathematics, University of California, Santa Barbara, CA 93106, USA}

%\maketitle

\begin{abstract}
We propose a deep learning algorithm for seismic interface and pocket detection with neural networks trained by synthetic high-frequency displacement data efficiently generated by the frozen Gaussian approximation (FGA). In seismic imaging high-frequency data is advantageous since it can provide high resolution of substructures. However, generation of sufficient synthetic high-frequency data sets for training neural networks is computationally challenging. This bottleneck is overcome by a highly scalable computational platform built upon the FGA, which comes from the semiclassical theory and approximates the wavefields by a sum of fixed-width (frozen) Gaussian wave packets.

Training data for deep neural networks is generated from a forward simulation of the elastic wave equation using the FGA.  This data contains accurate traveltime information (from the ray path) but not exact amplitude information (with asymptotic errors not shrinking to zero even at extremely fine numerical resolution). Using this data we build convolutional neural network models using an open source API, GeoSeg, developed using Keras and Tensorflow. On a simple model, networks, despite only being trained on data generated by the FGA, can detect an interface with a high success rate from displacement data generated by the spectral element method. Benchmark tests are done for P-waves (acoustic) and P- and S-waves (elastic) generated using the FGA and a spectral element method. Further, results with a high accuracy are shown for more complicated geometries including a three-layered model, a sine interface, and a 2D-pocket model where the neural networks are trained by both clean and noisy data.
\end{abstract}


\maketitle

%---CONTENT------------------------------------------------------------------------------
\section{Introduction}\label{sec:1}
Various geophysical aspects, e.g., tectonics and geodynamics~\citep{Aki1976,Romanowicz1991,Rawlinson2010,Zhao2012a}, can be better understood by images of substructures (e.g. locations of seismic interfaces) of the Earth generated by seismic tomography. Neural networks excel at recognizing shapes, patterns, and sorting relevant from irrelevant data; this makes them good for image recognition and classification. In particular, convolutional neural networks allowed for rapid advances in image classification and object detection~\citep{DBLP:journals/nature/LeCunBH15}, and in fact networks have been created for specific tasks, such as, fault detection~\citep{doi:10.1190/tle36030208.1}, earthquake detection, \emph{ConvNetQuake}~\citep{Perole1700578}, \emph{DeepDetect}~\citep{8424545} and seismic phase arrival times, \emph{PhaseNet}~\citep{2018arXiv180303211Z}.
One obstacle in building a neural network to detect seismic structures is having an ample data set for training. There is constant waveform data being collected by seismic stations across the globe, and generating data by resampling of this seismic data to train a network can be done, but is limited by the Nyquist frequency. Seismic data can not be resampled with a Nyquist frequency lower than the highest usable frequency in the data, thus high frequency data is usually preferred as it tends to lead to improved resolution of the substructures. Other difficulties of gathering an ample data lie within the differences in geological locations, natural phenomenon (e.g. earthquakes) and unnatural phenomenon (e.g. fracking). Using these data sets to train a general neural network is a daunting task, and thus it is natural to use synthetic data for the training of neural networks.

The dominant frequency of a typical earthquake is around 5 Hz~\citep{nakamichi2003source} leading to demanding, and at times, unaffordable computational cost. This makes generation of sufficient synthetic high-frequency data sets for training neural networks computationally challenging to well-known methods. We overcome this difficulty by building a highly scalable computational platform upon the frozen Gaussian approximation (FGA) method for the elastic wave equation \citep{hateley:2018}, which comes from the semiclassical theory. The FGA approximates the wavefields by a sum of fixed-width (frozen) Gaussian wave packets. The dynamics of each Gaussian wave packet follow ray paths with the prefactor amplitude equation derived from an asymptotic expansion on the phase plane.  The whole set of governing equations are decoupled for each Gaussian wave packet, and thus, in theory, each corresponding ODE system can be solved on its own process, making the algorithm embarrassingly parallel.

Using synthetic data, Araya-Polo et al. perform inverse tomography via fully connected neural networks with great success  in {\cite{araya2018deep}} . Their networks use low dimensional features extracted from seismic data as input. Using deeper convolutional neural networks trained on seismogram data may allow the network to pick up on previously unknown signals. The increase in input dimensionality necessitates more sophisticated deep learning techniques than those presented in {\cite{araya2018deep}}. 

In this paper, we propose a deep learning algorithm for seismic interface detection, with the neural networks trained by synthetic high-frequency seismograms. We first generate the time series of synthetic seismogram data by the FGA, which we use to train neural networks made with 
%, with which, we build network models using 
an open source API, GeoSeg, developed using Keras and Tensorflow. Despite only being trained on FGA generated data we observe the networks are able to detect a 1D interface with a high success rate on data generated by spectral element method. This method more acucurately represents true seismic signals when fine time step and mesh sizes are used in the computation. We conjecture that this robustness is due to the fact that although FGA does not carry exact amplitude information (with asymptotic errors proportional to the ratio of wavelength over domain size), it contains accurate traveltime information. For this simple problem it is straight-forward in geophysics to identify the traveltime as a key factor in interface location; however, this is not built into the network and so its use must be learned. With the success of the 1D interface detection, we further apply the deep learning algorithms for geometries with more complicated structures, including a three layered model and a 2D pocket model, both of which show a high accuracy. We also investigate the effect of noise by studying the performance of deep learning algorithms on noisy validation data, with the neural networks trained using clean and noisy data, respectively.
%We observe that networks, despite only being trained on FGA data, can detect an interface with a high success rate from the seismograms generated by the spectral element method, which more accurately represents true seismic signals when fine time step and mesh size are used in the computation. This is due to the fact that, although FGA does not carry exact amplitude information (with asymptotic errors proportional to the ratio of wavelength over domain size), it contains accurate traveltime information. 
%We present the performance by considering a simple two-dimensional layered velocity model, with synthetic data generated for both P- and S-waves. 
%We remark that, although it is straightforward in geophysics to identify traveltime as a key factor to detect the interface location for this example, it is not always natural for the trained networks to automatically use this physical intuition.

The paper is outlined as follows: In Section~\ref{sec:2}, we review briefly the mathematical background of FGA and describe how the synthetic data is generated. In Section~\ref{sec:network_design}, we describe the details of the network design including network and block architectures. In Section~\ref{sec:experiments} we show the performance of various networks on a series of geometries with different substructures, using both clean and noisy data. Concluding remarks are made in Section~\ref{sec:conclusion}.
%---Background--------------------------------------------------------------------------------
\section{Frozen Gaussian approximation}
\label{sec:2}
%---Background--------------------------------------------------------------------------------
We summarize the mathematical theory of FGA in this section; for full exposition and details for the elastic wave equation, see~\cite{hateley:2018}; and for the acoustic wave equation, see \cite{chai2017frozen}.  

The core idea of the FGA is to approximate seismic wavefields by fixed-width Gaussian wave packets whose dynamics follow ray paths with the prefactor amplitude equation derived from an asymptotic expansion on the phase plane.  The ODE system governing the dynamics for each wave packet are decoupled. In theory, each ODE system can be solved on its own process, hence it is embarrassingly parallel.  The implementation, as in previous works~\citep{hateley:2018}, is with Fortran using message passage interface (MPI). The implementation has a speed up factor of approximately 1.94; hence, doubling the number of cores nearly halves the computational time.
The equation for the forward modeling to generate the training data set we use is the elastic wave equation~\citep{dziewonski1981preliminary},
\begin{equation}\label{eq:1}
\rho\partial^2_t\bu = (\lambda + \mu)\nabla(\nabla\cdot \bu ) + \mu\Delta\bu + \bF,
\end{equation}
where $\rho,\lambda,\mu,:\mR^3\to\mR$ is the material density, the first and second Lam\'e parameters respectively and $\bu:\mR\times\mR^3\to\mR^3$ is displacement.  The differential operators are taken in terms of the spacial variables. Eq.~\eqref{eq:1} has a natural separation into divergence and curl free components and can also be written as
\begin{equation}\label{eq:2}
\partial^2_t\bu = c_\tp^2\nabla(\nabla\cdot \bu) - c_\ts^2\nabla\times\nabla \times \bu + \bF_\rho.
\end{equation}
This decomposition represents P-wave, and S-wave respectively with velocities
\begin{equation}\label{eq:3}
c^2_\tp(\bx) = \frac{\lambda(\bx) + 2\mu(\bx)}{\rho(\bx)},\indent c^2_\ts(\bx) = \frac{\mu(\bx)}{\rho(\bx)},
\end{equation}
with $c_\tp(\bx)$ representing the P-wave speed and $c_\ts(\bx)$ representing the S-wave speed.
\subsection{The FGA Formulation}
%-------------------
Presented below is an outline for the FGA.  For derivation and benchmarking tests we refer to~\citep{hateley:2018, chai2017frozen}.  We introduce the FGA formula for the elastic wave equation~\eqref{eq:2}, with initial conditions
\begin{equation}\label{eq:initcond}
\begin{cases}
&\bu(0,\bx)   = \bff^k(\bx), \\
&\partial_t\bu(0,\bx) = \bg^k(\bx), \\
\end{cases}
\end{equation}
where the superscript $k$ represents the wavenumber. For a sake of simplicity and clarity, we shall also use the following notations:
\begin{itemize}
\item $\I=\sqrt{-1}$ : the imaginary unit;
\item subscripts/superscripts ``$\tp$'' and ``$\ts$'' indicate P- and S-waves, respectively; 
\item $\pm$ indicates the two-way wave propagation directions correspondingly;
\item $\bd{\hat{N}}_{\tp,\ts}(t)$: unit vectors indicating the polarized directions of P- and S-waves;
\item  $\bd{\hat{n}}_{\tp,\ts}$: the initial directions of P- and S-waves.
\end{itemize}
 The FGA approximates the wavefield $\bu^k(t,\bx)$ in eq.~\eqref{eq:1} by a summation of dynamic frozen Gaussian wave packets,
  \begin{equation}\label{eq:FGA}
  \begin{aligned}
     u^{k}_{\FGA}(t, \bd{x}) & \approx \sum_{(\bd{q},\bd{p})\in G_\pm^\tp} \frac{a_\tp\bd{\hat{N}}_\tp
      \psi^k_\tp}{(2\pi/k)^{9/2}}
    e^{{\I}{k}\bd{P}_\tp\cdot(\bd{x} - \bd{Q}_\tp) -
      \frac{k}{2} \abs{\bd{x}
        - \bd{Q}_\tp}^2}\delta\bd{q}\delta \bd{p}  \\
    & + \sum_{(\bd{q},\bd{p})\in G_\pm^\ts} \frac{a_\ts\bd{\hat{N}}_\ts\psi^k_\ts}{(2\pi /k)^{9/2}}
    e^{{\I}{k} \bd{P}_\ts\cdot(\bd{x} - \bd{Q}_\ts) -
      \frac{k}{2} \abs{\bd{x} - \bd{Q}_\ts}^2}\delta\bd{q}\delta \bd{p},
  \end{aligned}
\end{equation}
with the weight functions
\begin{align}\label{eq:psi}
&  \psi^k_{\tp,\ts}(\bd{q}, \bd{p}) =
   \int \alpha^k_{\tp,\ts}(\bd{y},\bd{q},\bd{p})
  e^{- \I k \bd{p}\cdot(\bd{y} - \bd{q}) -
    \frac{k}{2} \abs{\bd{y} - \bd{q}}^2} \ud\bd{y},\\ \label{eq:alpha^k}
&\alpha^k_{\tp,\ts}(\bd{y},\bd{q},\bd{p})=\frac{1}{2kc_{\tp,\ts}\abs{\bp}^3}\bigl(k\bd{f}^k(\bd{y})c_{\tp,\ts}\abs{\bd{p}}\pm
  \I\bd{g}^k(\bd{y}) \bigr)\cdot\bd{\hat{n}}_{\tp,\ts}.
\end{align}
In eq.~\eqref{eq:FGA}, $G_\pm^{\tp,\ts}$ refers to the initial sets of Gaussian center $\bd{q}$ and propagation vector $\bd{p}$ for P- and S-waves, respectively.  In eq.~\eqref{eq:alpha^k}, the ``$\pm$'' on the right-hand-side of the equation indicate that the $\alpha^k_{\tp,\ts}$ correspond to $(\bd{q},\bd{p})\in G_\pm^{\tp,\ts}$. We refer \cite{hateley:2018} for the derivation, accuracy and explanation of FGA, and only summarize the formulation as follows. 
The ray path is given by the Hamiltonian system with Hamiltonian $H(\bd{Q},\bd{P})=\pm c_{\tp,\ts}(\bd{Q})\abs{\bd{P}}$.  The ``$\pm$'' give the two-way wave propagation directions; e.g. for the ``+'' wave propagation, $(\bd{q},\bd{p})\in G_+^{\tp,\ts}$, the Gaussian center $\bd{Q}_{\tp,\ts}(t,\bd{q},\bd{p})$ and propagation vector $\bd{P}_{\tp,\ts}(t,\bd{q},\bd{p})$ follow the ray dynamics
\begin{equation}\label{eq:char_plus}
  \begin{cases}
    \displaystyle
    \frac{\ud \bd{Q}_{\tp,\ts}}{\ud t} = c_{\tp,\ts}(\bd{Q}_{\tp,\ts})
\frac{\bd{P}_{\tp,\ts}}{\abs{\bd{P}_{\tp,\ts}}},\\[.5em]
    \displaystyle
    \frac{\ud \bd{P}_{\tp,\ts}}{\ud t} =- \partial_{\bd{Q}} c_{\tp,\ts}(\bd{Q}_{\tp,\ts})
\abs{\bd{P}_{\tp,\ts}},
  \end{cases}
\end{equation}
with initial conditions
\begin{equation}\label{eq:ini_plus}
\bd{Q}_{\tp,\ts}(0, \bd{q}, \bd{p}) =
\bd{q} \quad \text{and} \quad
\bd{P}_{\tp,\ts}(0, \bd{q}, \bd{p}) = \bd{p}.
\end{equation}
The prefactor amplitudes $\bd{a}_{\tp,\ts}(t,\bd{q},\bd{p})$ satisfy the following equations, where S-waves have been decomposed into SH- and SV-waves,
\begin{align}\label{eq:amp_P}
&\frac{\ud a_{\tp}}{\ud t} =  a_{\tp}\biggl(\pm\frac{\partial_{\bd{Q}_{\tp}}c_{\tp}\cdot \bd{P}_{\tp}}{|\bd{P}_{\tp}|} + \frac{1}{2}\TR\bigl(Z_{\tp}^{-1}\frac{\ud Z_{\tp}}{\ud t}\bigr)\biggr), \\
& \frac{\ud a_{\tsv}}{\ud t} =  a_{\tsv}\biggl(\pm\frac{\partial_{\bd{Q}_{\ts}}c_{\ts}\cdot \bd{P}_{\ts}}{|\bd{P}_{\ts}|} + \frac{1}{2}\TR\bigl(Z_{\ts}^{-1}\frac{\ud Z_{\ts}}{\ud t}\bigr)\biggr)-a_{\tsh}\frac{\ud \bd{\hat{N}}_{\tsh}}{\ud t}\cdot\bd{\hat{N}}_\tsv, \label{eq:amp_SV}\\
& \frac{\ud a_{\tsh}}{\ud t} = a_{\tsh}\biggl(\pm\frac{\partial_{\bd{Q}_{\ts}}c_{\ts}\cdot \bd{P}_{\ts}}{|\bd{P}_{\ts}|} + \frac{1}{2}\TR\bigl(Z_{\ts}^{-1}\frac{\ud Z_{\ts}}{\ud t}\bigr)\biggr)+a_{\tsv}\frac{\ud \bd{\hat{N}}_{\tsh}}{\ud t}\cdot\bd{\hat{N}}_\tsv, \label{eq:amp_SH}
\end{align}
with the initial conditions $a_{\tp,\tsv,\tsh}=2^{3/2}$, and $\bd{\hat{N}}_{\tsv}$ and $\bd{\hat{N}}_{\tsh}$ are the two unit directions perpendicular to $\bd{P}_\ts$, referring to the polarized directions of SV- and SH-waves, respectively. With the short-hand notations,
 \begin{equation}\label{eq:op_zZ_app}
  \partial_{\bd{z}}=\partial_{\bd{q}}-\I\partial_{\bd{p}},
  \qquad
  Z_{\tp,\ts}=\partial_{\bd{z}}(\bd{Q}_{\tp,\ts}+\I\bd{P}_{\tp,\ts%%
 }).
\end{equation}

\begin{figure}
	\centering
	\includegraphics[scale=0.4]{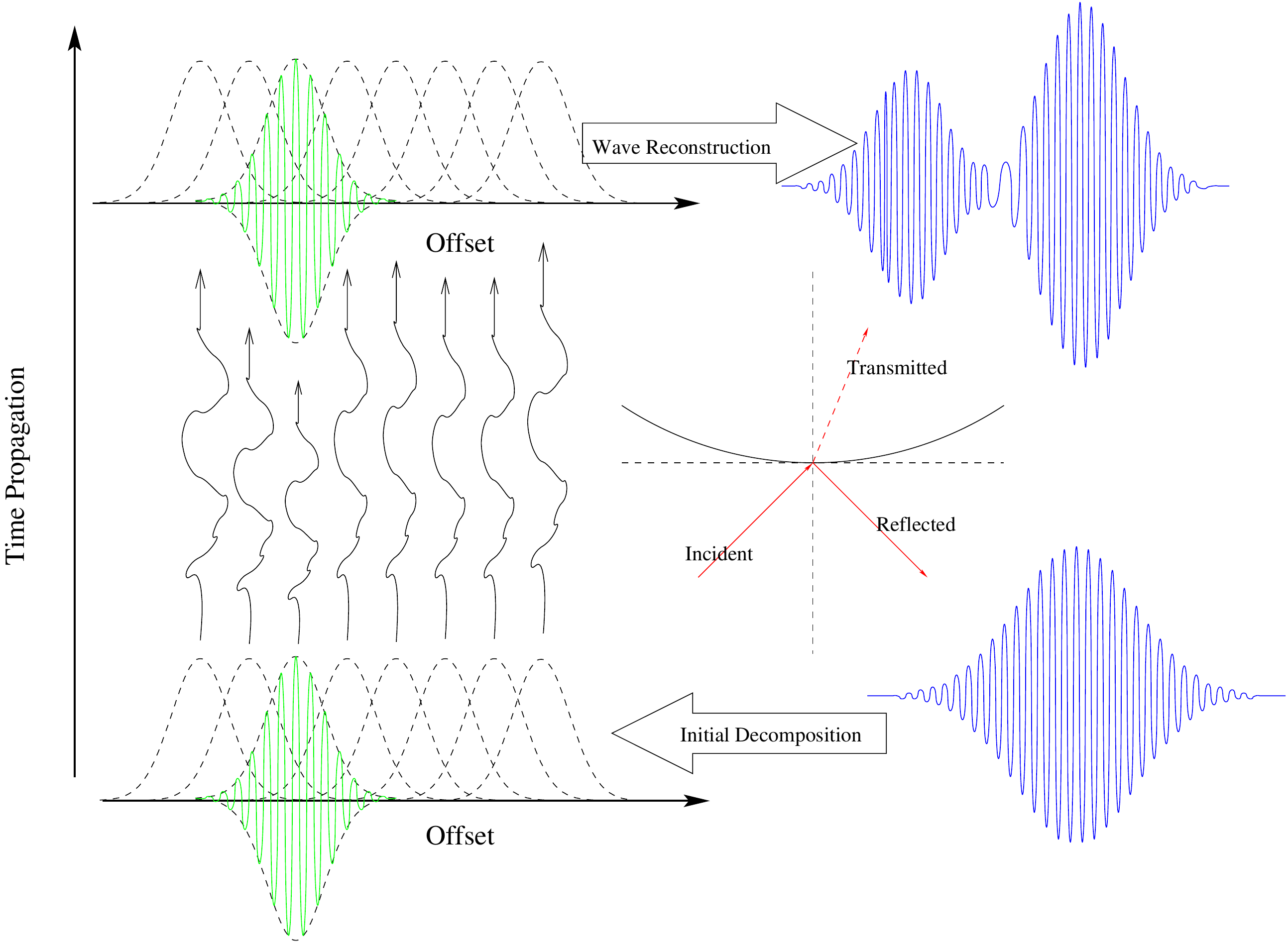}
	\caption{{A cartoon illustration of FGA algorithms: Step 1, decompose the initial wavefield into a sum of Gaussian wave packets with corresponding weights given by \eqref{eq:psi}; Step 2, propagate Gaussian wave packets following \eqref{eq:char_plus}, \eqref{eq:amp_P}, \eqref{eq:amp_SV} and \eqref{eq:amp_SH}, with the reflection-transmission conditions described in Section~\ref{sec:interface}; Step 3, reconstruct the wavefield by summing all Gaussian wave packets using \eqref{eq:FGA}. } }\label{fig:algm}
		\end{figure}

We illustrate the algorithm by Figure~\ref{fig:algm}, and refer to the Figures~5 and 6 in \cite{hateley:2018} for the performance of efficiency of FGA.

\subsection{Interface conditions}\label{sec:interface}
\begin{figure}
\begin{center}
\input{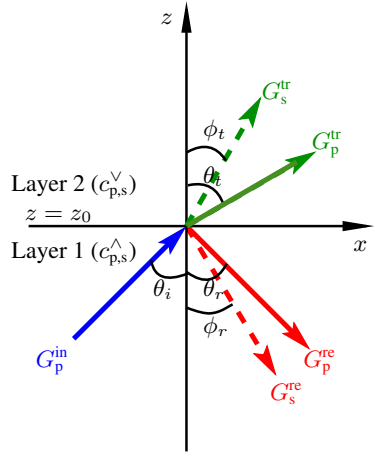}
\end{center}
\caption{Cartoon illustration of an incident Gaussian wave packet for P-wave hitting the interface at $z = z_0$, and then reflected and transmitted as Gaussian wave packets for P- and SV-waves. Here the $G^{\text{in,re,tr}}_{\text{p,s}}$ stands for the Gaussian wave packet for the incident, reflected and transmitted P- and SV-waves, respectively. We denote $\theta_i, \theta_r, \theta_t$ to be the incident, reflection and transmission angles of P-waves, and $\phi_r, \phi_t$ to be the reflection and transmission angles of SV-waves, respectively.}\label{fig:interface}
\end{figure}

Interface conditions are important as the direct and reflected waves from an interface are picked up by the receiver, which records the time series of wavefield at certain location.  This gives travel time information; which in turn enables the depth of an interface to be computed. For this exposition we only consider a flat interface, in general, we can use tangential-normal coordinates. The derivation was detailed in Appendix~B in \cite{hateley:2018}, with the idea of using the continuity of level set functions corresponding to the Hamiltonian dynamics \eqref{eq:char_plus}. A cartoon illustration on the behavior of Gaussian wave packet is given in Figure~\ref{fig:interface}.  For a flat interface $z = z_0$, the wave speeds of the two layers near the interface are assumed to be, 
\begin{align}
c_{\tp}(\bd{x}) =
\begin{cases}
c_{\tp}^\vee(\bd{x}) & z > z_0 \\
c_{\tp}^\wedge(\bd{x}) & z < z_0
\end{cases},\quad
c_{\ts}(\bd{x}) =
\begin{cases}
c_{\ts}^\vee(\bd{x}) & z > z_0 \\
c_{\ts}^\wedge(\bd{x}) & z < z_0
\end{cases}.
\end{align}
As a Gaussian wave packet hits an interface, several of its quantities need to be defined. First, $a_{\tp,\ts}$ and $\bd{P}_{\tp,\ts}$, are determined by Snell's Law and the Zoeppritz equations~\citep{Yi:01}. If one denotes $\theta_i, \theta_r, \theta_t$ to be the P-wave incident, reflection and transmission angles, and $\phi_r, \phi_t$ to be the SV-wave reflection and transmission angles, respectively, then the Zoeppritz equations read as 
\begin{equation}\label{eq:Yilmaz}
M\left(
\begin{array}{c}
a_\tp^{\RE} \\
a_\ts^{\RE} \\
a_\tp^{\TR} \\
a_\ts^{\TR} \\
\end{array}
\right)
=
\left(
\begin{array}{c}
\cos(\theta_r) \\
\sin(\theta_r) \\
\cos(2\phi_r) \\
\cos(2\theta_r) \\
\end{array}
\right)a_\tp^{\IN},
\end{equation}
with the matrix M as
\begin{equation}
 M =
\left(\begin{array}{c c c c}
\cos(\theta_r) & \frac{c_{\tp}^\wedge}{c_{\ts}^\wedge}\sin(\phi_r)      & \frac{c_{\tp}^\wedge}{c_{\tp}^\vee}\cos(\theta_t) & -\frac{c_{\tp}^\wedge}{c_{\ts}^\vee}\sin(\phi_t) \\
-\sin(\theta_r) & \frac{c_{\tp}^\wedge}{c_{\ts}^\wedge}\cos(\phi_r)     & \frac{c_{\tp}^\wedge}{c_{\tp}^\vee}\sin(\theta_t) & \frac{c_{\tp}^\wedge}{c_{\ts}^\vee}\cos(\phi_t) \\
-\cos(2\phi_r) & -\sin(2\phi_r)                      & \frac{\rho_2}{\rho_1}\cos(2\phi_t) & -\frac{\rho_2}{\rho_1}\sin(2\phi_t) \\
\sin(2\theta_r) & -(\frac{c_{\tp}^\wedge}{c_{\ts}^\wedge})^2\cos(2\phi_r)& \frac{\rho_2(c_{\tp}^\wedge c_{\ts}^\vee)^2}{\rho_1(c_{\tp}^\vee c_{\ts}^\wedge)^2}\sin(2\theta_t) & \frac{\rho_2(c_{\tp}^\wedge)^2}{\rho_1(c_{\ts}^\wedge)^2}\cos(2\phi_t)
\end{array}
\right),
\end{equation}
where $\rho_{1,2}$ are the densities for the layers 1 and 2, respectively. Let $\bN$ denote the normal to the interface at the point of incidence then $\bQ^{\IN,\RE,\TR}$ is the Gaussian center at the point of incidence, and $\bP^{\IN,\RE,\TR}$ corresponds to the propagation vector of incident, reflected and transmitted Gaussian wave packet for either P- or S-waves. $\bQ^{\IN} = \bQ^{\RE} = \bQ^{\TR}$ and $\bP^{\RE,\TR}$ is updated as follows
\begin{align} \label{eq:p:int} 
\bP^{\TR,\RE}_{\tp,\ts}&=\bP^{\IN} + \sgn(\bP^{\TR,\RE}_{\tp,\ts})\Big(\sqrt{|\bP^\IN|n_{\tp,\ts}^{\TR,\RE} - \big| |\bP^\IN| -  (\bP^\IN\cdot\bN)^2 \big|} - (\bP\cdot \bN)\Big)\bN, \\ \nonumber
\end{align}
where $n_{\tp,\ts}^{\TR,\RE}$ denotes the index of refraction for the new respective direction, e.g. $n_{\tp}^{\TR} = c_{\tp}^\vee/ c_{\tp}^\wedge$. Also $Z_{\tp,\ts}$ needs to be updated, requiring use of conservation of level set functions defined in the Eulerian frozen Gaussian approximation formula \citep{LuYa:MMS,WeYa:12}.  
\begin{equation}\label{eq:Inter_DzQDzP}
\begin{aligned}
    \p_\bz\bQ^{\RE,\TR}&=\p_\bz\bQ^{\IN}\,F,\\
    \p_\bz\bP^{\RE,\TR}&=\p_\bz\bP^{\IN}\,W-\f{\abs{\bP^{\RE,\TR}}}{c(\bQ^{\RE,\TR})\bP^{\RE,\TR}\cdot \bN}\left(\p_\bz\bQ^{\RE,\TR}
    \cdot{\nb}c(\bQ^{\RE,\TR})-\p_\bz\bQ^{\IN}\cdot{\nb}c(\bQ^{\IN})\right)\bN,
\end{aligned}
\end{equation}
$F$ and $W$ are two $3\times3$ matrices, $F^T=W^{-1}$, and
   \begin{align*}
    F=\begin{bmatrix}
    1& 0& 0\\0&1&0\\
    \left(\kappa-1\right)\f{p_x}{p_z^{\IN}}&\left(\kappa-1\right)\f{p_y}{p_z^{\IN}}&\kappa\f{p_z^{\RE,\TR}}{p_z^{\IN}}
      \end{bmatrix},\quad\text{with}\quad \kappa=\left(\frac{c(\bQ^{\RE,\TR})}{c(\bQ^{\IN})}\right)^2.
   \end{align*}

\subsection{Advantage of FGA for Generating Training Data}
 The data points used for our experiments are generated from the forward simulation of the elastic wave equation using the FGA. We record the displacement data from the wavefield at various points near the surface; these points represent the receiver locations.
Given an initial condition, as in eq.~\eqref{eq:initcond}, the initial wave packet decomposition can be saved for generating a data set for training.  
That is, the same data can be loaded as the parameters which vary from data point to data point; e.g. interface height, pocket location, pocket size, etc. Furthermore, if the initial condition is independent of the wave velocities, the same wave packet decomposition can be used to generate data from simulation with varying velocities.

For a single forward simulation; after the initial wave packet decomposition generated and saved, loading the initial wave packet decomposition, running an ODE solver, and recording the displacement are the only tasks required to generate a data point. For generation of a data set, the simulation can be restarted at $t=0$ with another set of parameters. As the initial wave packet decomposition is already loaded in memory,  all that is required to generate the rest of the data set is running an ODE solver, and recording the displacement. The ODE system for the FGA is uncoupled for each wave packet, the speed of a single simulation greatly benefits from a parallel implementation. 

\section{Network Design} \label{sec:network_design}
  
%\subsection{Generation of the Synthetic Data}
%Large and diverse datasets can reduce generalization error of a network. 

%A CMP stack can be thought of as an image from a seismic event in which information about subsurface structures is stored.  In terms of reflective seismology a seismogram records information and the refractive index of an interface is encoded in that seismogram. Seismic migration is a process by which seismic events are geometrically re-located to their incidence location, rather than the recorded location, in the subsurface. 

The goal of Full Waveform Inversion (FWI) is to extract wave speed data from seismic data. In its purest form, this is a regression type problem and was addressed with fully connected networks in {\cite{araya2018deep}}. Our work approaches the problem from a segmentation perspective. We address a simplified version of FWI and attempt to detect subsurface structures by classifying them as regions of low or high wavespeed, thus transforming the regression problem into a segmentation problem. These sorts of segmentation problems have been addressed with great success by CNNs \citep{Shelhamer2015FullyCN}. 
Semantic segmentation of images is the process of labeling each pixel in an image with a class label for which it belongs. In semantic segmentation problems the correct pixel label map is referred to as the ground truth.
	In our work the ``image” is the n-dimensional slice in the depth direction which is partitioned into N bins. The $i,j^{th}$ ``pixel'' is the signal value from receiver $i$ at depth bin $j$.

 Each bin is then labeled depending on whether it came from a region of high or low velocity. These velocity regions are our classes. Our work diverges substantially from traditional semantic segmentation of images, as our input is time series data which must be transformed by the network.  This is opposed to the traditional case where the input itself is labeled. 
%The goal of our network is given a time series of seismogram data to infer the presence of high and low wave speed regions and the interfaces between them. The input of all the neural networks will be seismic data read from three receivers and the corresponding ground truth for a sample will be a vector, whose index represents a normalized depth for the simulation, and whose values are either 0 or 1, representing a high or low wave speed material present at that depth represented by the index. We remark that the number of receivers is not a limitation for either the generation of the data nor the training of the neural network.
The goal of our network is to infer the presence of high and low wavespeed regions and the interfaces between them from seismogram data. The input to the network is $X \in \mathbb{R}^{M\times d \times r}$, where $M$ is the number of timesteps, $d$ is the spatial dimension of media, and $r$ is the the number of receiver. The output of the network is 
\begin{equation}
\mathcal{N}(X) = (p_{i_1\dots i_n}^k) \in \mathbb{R}^{M_1 \times \cdots \times M_n \times N}, \quad \begin{array}{l} i_j \in \{1, ... ,M_j\} \\ k \in \{1,\dots, N\} \end{array},
\end{equation}
where $p_{i_1\cdots i_n}^k$ is the probability that bin $i_1\cdots i_n$ belongs to the $k^{th}$ class. In this paper $d=3$, $n=1,2$, and $N=1,2,3$.

For example, possible output and groundtruth could be
%For each sample, the output of the neural network is a real valued matrix whose rows represent depth, and whose columns represent the probability of a given row belonging to either the first material or the second. For example, consider the output matrix and ground truth for a fixed sample:
\begin{align*}
\begin{bmatrix}
	0.1& 0.9\\
	0.2& 0.8\\
	0.55& 0.45
\end{bmatrix}, \quad
\begin{bmatrix}
1\\
1\\
1
\end{bmatrix}.
\end{align*}
Here, at depth indexed by 1, the network believes with 10\% probability that this bin is a low speed region and with 90\% probability that it is a high speed region, and similarly for the other rows.  The accuracy of a given inference is found by taking the argmax along the last axis of the output tensor and comparing against the groundtruth.  Taking a max along the last axis recovers the probability, interpreted as a confidence of the prediction. The above example has 66.67\% accuracy, and the confidence is $[0.9, 0.8, 0.55]^{\text{T}}$.

%The training loss for the neural network is categorical cross-entropy. 
%Generically a model is composed of an input layer, \( N \) layers of selected blocks following the selected architecture, and then an output layer for classification.
In {\cite{araya2018deep}}, Araya-Polo et al. perform inverse tomography via Deep Learning and achieve impressive results. Our model is fundamentally different than GeoDNN in that: GeoDNN is a fully connected network whereas GeoSeg's is fully convolutional, and GeoDNN uses semblance panels from CMP data as features for the network and GeoSeg uses the raw seismograph data. Moreover, Araya-Polo et al. address the FWI problem and provide the wave speeds in a two dimensional region and we tackle high and low velocity detection, shifting the problem from regression to segmentation. 
%\subsection{Network Architectures}
%\label{network_arch}

The networks were built using an open source API, GeoSeg$^{\footnotemark}$\footnotetext{https://github.com/KyleMylonakis/GeoSeg}, developed using Keras and Tensorflow. GeoSeg supports UNet, fully convolutional segmentation network, or feed forward CNNs as a base meta-architecture, using any of residual, dense, or convolutional blocks, with or without batch normalization~\citep{Ronn_Fischer_UNet,Shelhamer2015FullyCN,he2016deep,Huang_Liu_Weinberger_DN,Ioffe_Szegedy_BN}. GeoSeg also allows for easy hyper-parameter selection for network and block architectures, and for training optimizers and parameters. The optimizers used were NADAM with default parameters~\citep{dozat2016incorporating}, sometimes followed by minibatch stochastic gradient descent (SGD), or SGD alone. 
The network structures are described by their meta-architecture and their blocks. The meta-architecture describes the global topology of the network and how the blocks interact with each-other. Each block either begins or ends with a decoding or encoding transition layer respectively. Encoding transition layers downsample their inputs with a strided convolution. Decoding transition layers upsample thier inputs with a strided deconvolution. Tranistion layers will not have dropout. \\
%that will down or up sample the temporal axis respectively. The block itself preserves the dimensions of its input, and never has dropout.

{\bf Meta-Architectures.} 
While GeoSeg supports many kinds of feed-forward CNN's and Encoder-Decoder Networks with different choices of blocks, UNet architectures with dense blocks performed the best and will be the only type of network reported. 

\begin{figure} [h t p]
  %\subfigure[GeoDUDe-2]
  %{\begin{minipage}[t]{0.5\linewidth}
  %\centering
  %\includegraphics*[scale = .5]{diag_unet.png}
  %\end{minipage}}
	%\subfigure[GeoDSegDe-2: with transfer branch \colorA{NEED TO CREATE THIS PICTURE}]
  %{\begin{minipage}[t]{0.5\linewidth}
  \centering
  \includegraphics*[scale = .8]{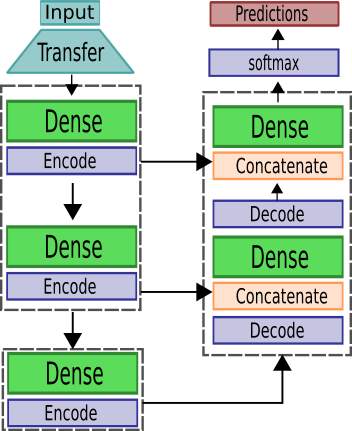}
  %\end{minipage}}
  \caption{Meta-architecture of a two-layered UNet, GeoDUDe-2, with Transfer Branch used in deep learning algorithms. For 2D problems the input is upsampled along the receiver axis by deconvolutions in the Transfer Branch. UNet's have ``rungs'' that connects the encoder and decoder branches. In this way, the network can incorporate both low and high resolution data. }\label{fig:meta_archs}
\end{figure}

GeoDUDe-L refers to a UNet architecture from~{\cite{Ronn_Fischer_UNet}}. These architectures have proven highly efficient at image segmentation for road detection~\citep{Zhang_Liu_RUNet} and in biomedical applications~\citep{Ronn_Fischer_UNet}. These networks feed their input into a transfer branch, then an encoder branch of length L, bridge block, and then a decoder branch of length L. The last layer is a convolutional layer followed by a softmax which outputs predictions as described above. The defining feature of these networks are the ``rungs'' connecting the encoder and decoder branches (see Figure~\ref{fig:meta_archs}). In this way, the network can incorporate both low and high resolution data~{\citep{Ronn_Fischer_UNet,Zhang_Liu_RUNet}}. For the one dimensional problems the transfer branch is not necessary and can be omitted. \\

\begin{figure} 
  \subfigure[Convolutional Layer]
  {\begin{minipage}[t]{0.5\linewidth}
  \centering
  \includegraphics*[scale = .8]{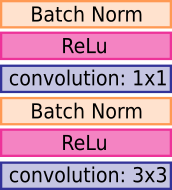}
  \end{minipage}}
  \subfigure[Dense Block]
  {\begin{minipage}[t]{0.5\linewidth}
  \centering
  \includegraphics*[scale = .8]{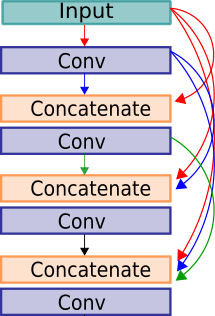}
  \end{minipage}}
  \caption{The type of blocks used in GeoSeg for this paper: (a) Block compositions of a basic convolutional layer using a bottleneck convolution to expand the filter channels before the full convolution; (b) a corresponding dense block. Each layer of the block recieves input from all previous layers allowing information to flow through the whole block.}\label{fig:blocks}
\end{figure}

{\bf Convolutional Layers.} The layer is broken first into a bottleneck convolution followed by the main convolution. The bottleneck is a convolution which uses a 1x1 kernel to expand the number of feature channels before performing the full convolution. It is suggested in {\cite{He_Zhang_Ren_Sun_ResNets,SpringenbergDBR14}} that such a bottleneck can reduce the number of necessary feature maps and so improve computational efficiency. We use Rectified Linear Units  (ReLUs) {\citep{Glorot2011DeepSR}} for our activation and size 3x3 (3x1 for 1D interface problems) filter kernels for our convolutions. As in {\cite{Huang_Liu_Weinberger_DN}}, we use Batch-Normalization {\citep{Ioffe_Szegedy_BN}} to help smooth training.  The setup is shown in Figure~\ref{fig:blocks} \\
%Where our layers depart from {\cite{Huang_Liu_Weinberger_DN}} is that our problem is one dimensional due to the piecewise constant assumption of the material wavespeeds. Because of this our final convolution is with kernel size 3x1. Thus ignoring spatial interactions until final predictions.

{\bf Dense Blocks.} Though GeoSeg supports multiple block types, all the networks reported in this paper use dense blocks. These are stacks of convolutional layers as shown in Figure~\ref{fig:blocks}. The defining features of these blocks, introduced in~{\cite{Huang_Liu_Weinberger_DN}} is that every layer receives input from all previous layers in the block via concatenation. Such architectures have been shown to greatly improve results in image classification while reducing computational burden~{\citep{Huang_Liu_Weinberger_DN}}.
%{\bf Naming Convention} We will use a naming convention that abbreviates meta-architecture, block type, and number of blocks per branch. GeoDSegDe-N will refer to deep encode/decode segmentation meta-architecture with N blocks each on the encode and decode branches, while GeoDUDe-N will refer to a UNet with meta-architecture using Dense blocks with N layers each on the encode and decode layers.
\\

{\bf Transfer Branch.} All of our meta-architectures preserve resolution of their input and so our detection resolution is limited by input resolution. This is not a problem in the temporal axis, which translates to the $z$ axis in output, since we have a large number of time samples; however, the $x$-axis resolution is limited by the number of receivers we have for our input. To increase the resolution in this direction, we place a small $l$-layer CNN before the main network which upsamples the receiver axis, via strided deconvolutions, by a factor of $2^l$.
\\

%{\bf Predictions.} As described above each pixel will be a probability of a region belonging to a given wave speed class. The final output of all meta-architectures presented will be a tensor of shape $M_1\times  \times 4$

\section{Numerical Experiments}
\label{sec:experiments}
%\section{Benchmark: FGA and SPE performance}\label{fga_spe_per}
Here we present the performance of deep learning algorithms for the three detection experiments: 1D interface problme, three-layered media model, and a 2D single cylindrical pocket model. The architecture used for all experiments is a UNet with Dense Blocks (GeoDUDe). Each dense block will be made of four constituent bottle-necked convolutional layers with a bottle neck factor of 4. For all 1D networks the dense blocks' convolutions use a kernel size of $3 \times 1$ in the base of the block and $2 \times 1$ at each transition layer, while for the 2D networks a $3 \times 3$ kernel size is used in the base block with a $2 \times 2$ kernel size in the transition layer. The meta-architectures had 16 filter channels except for the 1D interface model with P-wave data which only used 4. 

Our primary evaluation metric is accuracy which is the number of correctly predicted pixels over total pixels, {\it i.e.}, 
\begin{equation*}
\displaystyle\text{accuracy}=\frac{\text{number of correct pixels}}{\text{total number of pixels}},
\end{equation*}
where we set the ground truth as follows for measuring accuracy; if any part of a pixel contains a low velocity region, that pixel is counted as part of the low velocity region.

 For the 2D pocket model, we will also consider the Intersection Over Union metric which better captures segmentation performance. 

In the 2D pocket model, a two-layer transfer branch was used. Each layer was a convolution, two-strided in the receiver direction with a kernel size of $3 \times 3$ with 4 filter channels. During training, these layers had a drop out probability of 0.2.

The wavespeeds $c_{\tp}$ and $c_{\ts}$ are given by \eqref{eq:3}, which will be specified as piecewise linear functions (or constants) detailed in each numerical example. The initial P-Wave data is generated with source function
\begin{equation}\label{eq:srcfun}
f^k_j(\bx) = \cos(k(x_j - x_{0,j}))\exp\big(-2k|\bx - \bx_0|^2\big),
\end{equation}
and the P,S-Wave initial data is generated from the Green's function
\begin{equation}\label{eq:ex_anal}
\begin{aligned}
f_j^k(\bx) = \sum_{i=1}^3\ds\frac{(x_i-x_{i,0})(x_j-x_{j,0})}{4\pi\rho c_{\tp}^2r^3}F_j(t_0 - r/c_{\tp}) & +\\
\frac{r^2\delta_{ij} - (x_i-x_{i,0})(x_j-x_{j,0})}{4\pi\rho c_{\ts}^2r^3} & F_j(t_0 - r/c_{\ts}) + \\
\frac{3(x_i-x_{i,0})(x_j-x_{j,0}) - r^2\delta_{ij}}{4\pi\rho r^5} & \int_{r/c_{\tp}}^{r/c_{\ts}}sF_j(t_0 - s)\ \ud s,
\end{aligned}
\end{equation}
where $F_j(t) = \cos\left(kt\right)\exp(-2kt^2)$, $\delta_{ij}$ is the Kronecker delta, and $t_0 =2\sqrt{1/k}$, 

$\bx = (x_1,x_2, x_3)$, $\bx_0 = (x_{0,1},x_{0,2},x_{0,3})$ is the location of the source, $r = \|\bx - \bx_0\|$ and $\rho$ is the density. \\
The data is generated on the cluster, \emph{pod}, at the center for scientific computing at UC Santa Barbara$^{\footnotemark}$\footnotetext{http://csc.cnsi.ucsb.edu/clusters/pod} using 64 processes with a 4th order Runge-Kutta solver for the ODE system. As the initial condition is independent of the wavespeed, only one wave packet decomposition needs to be computed and saved for all data points to be generated.  This saves a tremendous amount of time as only the ODE system needs to be solved for various wavespeeds and interface heights. For example to generate the P-Wave data, when $804672$ total beams are used, each data point is generated in approximately $2.5$ minutes. This is compared to SPECFEM3D which takes is approximately $45$ minutes to generate a data point.

All of the networks were trained on the Google Cloud Platform, or on the cluster \emph{Pod} at the center for scientific computing at UC Santa Barbara with Keras 2.2.2 and Tensorflow 1.10.0 as a backend using a single NVIDIA Tesla V100 GPU.

%Twenty four neural network architectures were proposed by considering all choices of meta-architectures, blocks, in two and three block configurations with and without batch normalization. Each model was trained for \( 3500 \) epochs with an 80 percent split between training and evaluation data. For every dataset the high velocity region is always below the low velocity region. The training data set only contains time series data generated by the FGA. Separate models were trained for data sets containing only P-waves and those containing both P- and Ss waves. Models without batch normalization had poor evaluation accuracy and loss, and so will not be reported. All models were trained with dropout probability of 50\% {\cite{Srivastava2014DropoutAS}}.

\subsection{1D Interface}

To provide a proof of concept we first experimented with a two-layered flat interface model. We also use this case to investigate whether our network is simply inverting the FGA by comparing performance of a network trained on FGA but evaluated on data generated by SEM. 

\subsubsection{P-Wave Data}
\begin{figure} 
\subfigure[Computational Domain]
{\begin{minipage}[t]{0.5\linewidth}
\centering
\includegraphics*[scale = .35]{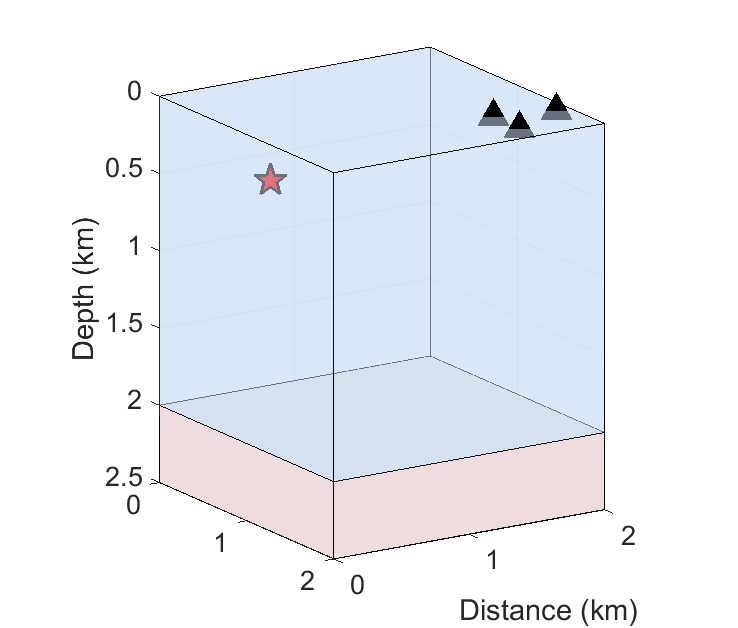}
\end{minipage}}%%
\subfigure[Seismogram]
{\begin{minipage}[t]{0.5\linewidth}
\centering
\includegraphics*[scale = .35]{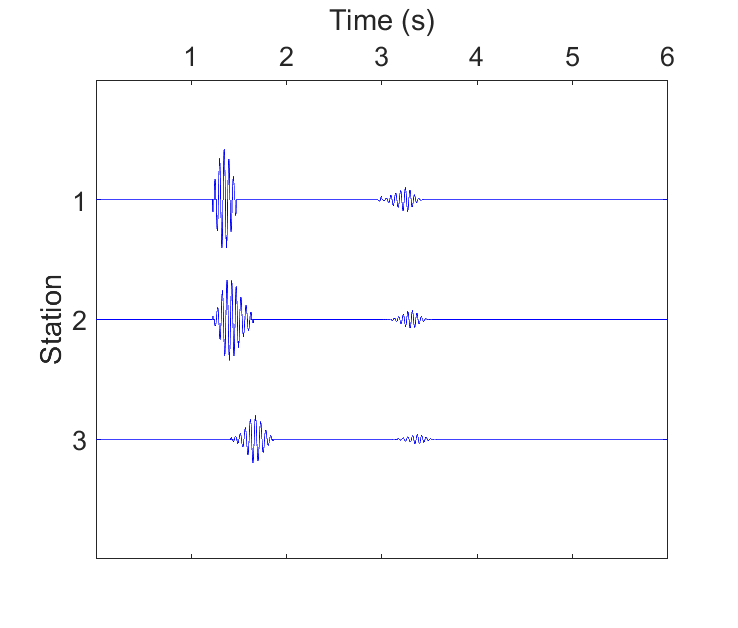}
\end{minipage}}
\caption{The locations of source and receivers, and the generated synthetic P-wave seismograms for the 1D interface problem. We take $k=128$ for generating the synthetic data. (a) The source is located at (.5,.5,.5) km as a star and the 3 receivers are located on the surface. The interface presented is at a depth of 2 km. (b) A visualization of typical data point, which is a collection of 3 seismograms from the forward simulation using the FGA.}\label{fig:p:setup}
\end{figure}

{\bf Dataset.} The P-wave data set is generated with a computation domain of $[0,2]$ km$\times[0,2]$ km$\times[0, 2.5]$ km with a source centered at $\bx_0 = (0.5, 0.5, 0.5)$ km and $k=128$ in \eqref{eq:srcfun}, which corresponds to approximately $20.37$ Hz. The stations are located on the surface at $S_1:(1.5,1.5,0)$ km, $S_2:(1.8,1.5,0)$ km, $S_3:(1.6,1.9,0)$ km. The interface is a plane, $z = z_0$ that varies from depth 1 km to 2.5 km. Above the interface the wavespeed $c_\tp$ varies linearly from .78 km/s to 1.22 km/s, below the interface the wavespeed $c_\tp$ varies linearly from 1.29 km/s to 1.56 km/s. See Figure~\ref{fig:p:setup}.

Each data point is a $(6000, 3, 3)$ tensor. Prior to training, we further down sample the temporal dimension by a factor of 25 and normalize the amplitude of the seismogram data. There were a total of $7790$ examples. The mini-batch size during training was $256$ examples.
\\

{\bf Network Details.} As described above our architecture was a 1D GeoDUDe-3 where each convolutional layer in the dense block had 4 feature channels. The  During training the dropout probability was set to 0.5 and a NADAM optimizer was used with default parameters. 
\\

{\bf Results.} Network evaluations were performed with data generated by the FGA and SPECFEM. Notably, the networks are never trained on any SPECFEM data. This was to investigate whether the network was sensitive to the asymptotic error produced by the FGA. 

%The training and evaluation loss and accuracy are shown in Fig.~\ref{fig:1}.

%%%%%%%%%%%%%%%%%%%%%%%%%%%%%%%%%%%%%%%%%%5
% 
%%%%%%%%%%%%%%%%%%%%%%%%%%%%%%%%%%%%%%%%%5
%\begin{figure} 
%\subfigure[Evaluation Accuracy ]
%{\begin{minipage}[t]{0.5\linewidth}
%\centering
%\includegraphics*[scale = .35]{P_eval_acc.png}
%\end{minipage}}%%
%\subfigure[Training Accuracy]
%{\begin{minipage}[t]{0.5\linewidth}
%\centering
%\includegraphics*[scale = .35]{P_train_acc.png}
%\end{minipage}}
%\caption{P-wave training results. The evaluation dataset for this figure only contains data generated by the FGA. Accuracy is computed as the total number correct classifications.}\label{fig:1}
%\end{figure}

%All of the final training and evaluation parameters are reported in Table~\ref{tab:table1}.

%\begin{table}[h!]
	%\begin{center}
		%\caption{P-Data Network}
		%\label{tab:table1}
		%\begin{tabular}{r|r|r|r|r|r}
			%\textbf{Network} & \textbf{Eval. Acc.} & \textbf{Eval. Loss} & %\textbf{Train. Acc.} & \textbf{Train. Loss} & \textbf{SEM Acc.}\\
%			\hline
%			GeoDUDe-3 & 96.97 \% & 0.09700 & 98.86 \% & 0.03388  & 94.29 \% \\
%		\end{tabular}
%	\end{center}
%\end{table}

\begin{figure} \label{fig:p_fga_results}
\subfigure[FGA: Actual]
{\begin{minipage}[t]{0.33\linewidth}
\centering
\includegraphics*[scale = .25]{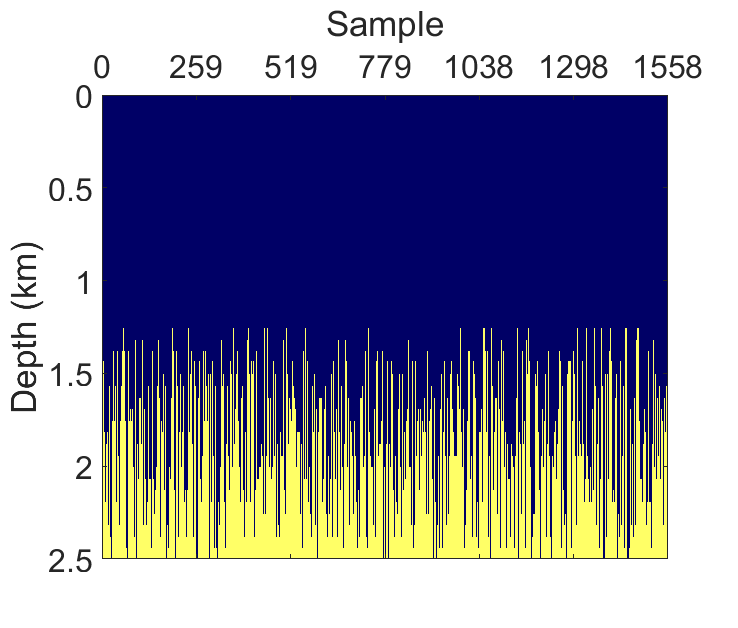}
\end{minipage}}%%
\subfigure[FGA: Predicted]
{\begin{minipage}[t]{0.33\linewidth}
\centering
\includegraphics*[scale = .25]{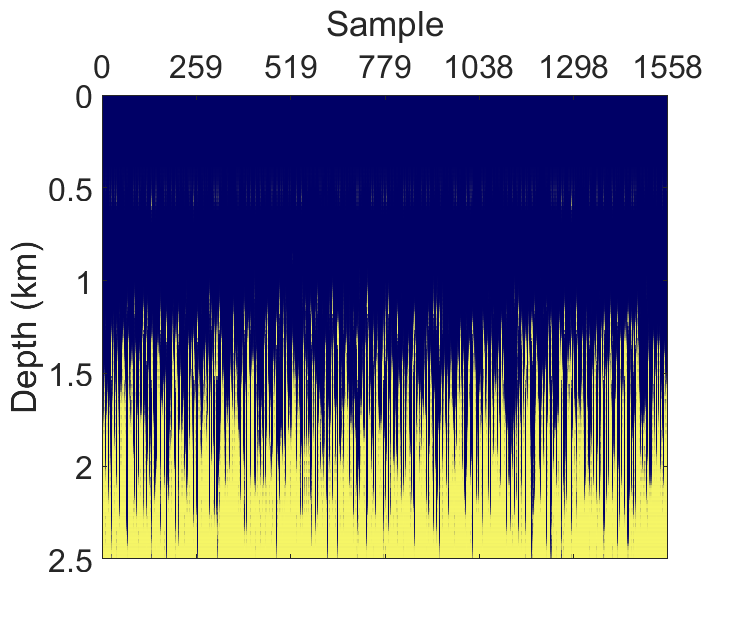}
\end{minipage}}
\subfigure[FGA: Difference]
{\begin{minipage}[t]{0.33\linewidth}
\centering
\includegraphics*[scale = .25]{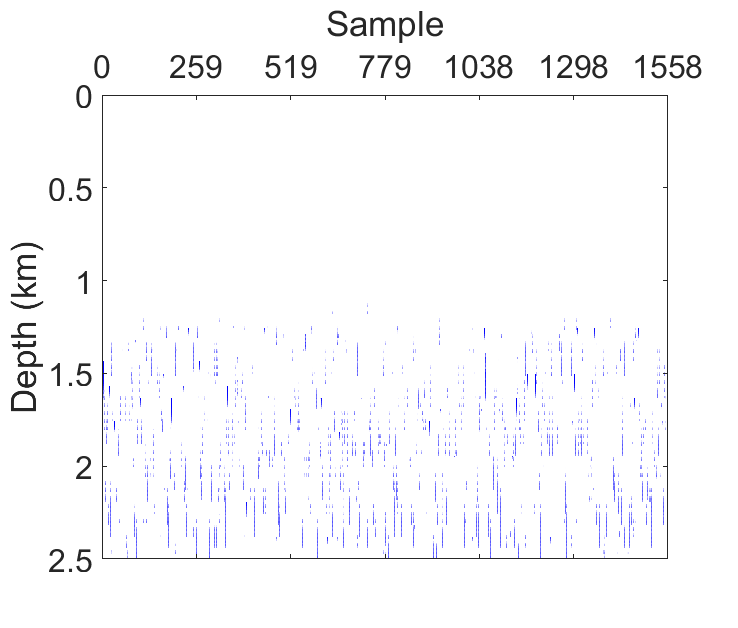}
\end{minipage}} \\
\subfigure[SEM: Actual]
{\begin{minipage}[t]{0.33\linewidth}
\centering
\includegraphics*[scale = .25]{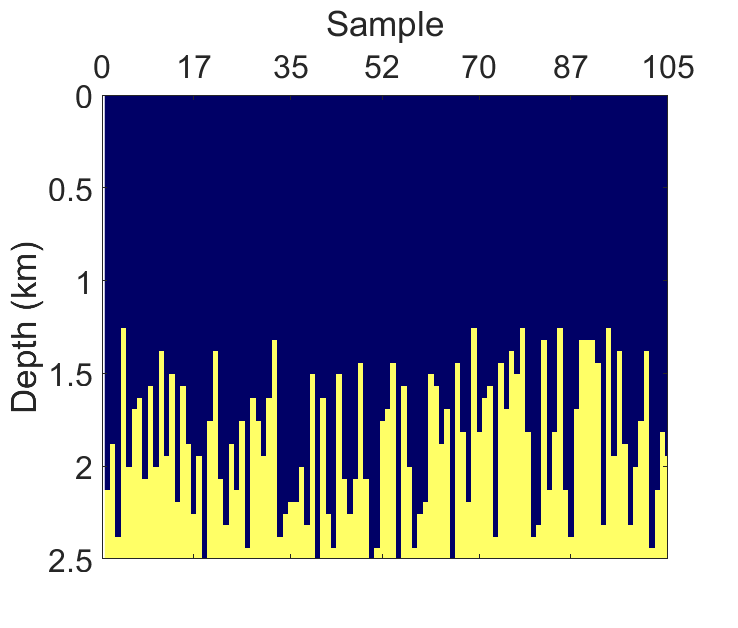}
\end{minipage}}%%
\subfigure[SEM: Predicted]
{\begin{minipage}[t]{0.33\linewidth}
\centering
\includegraphics*[scale = .25]{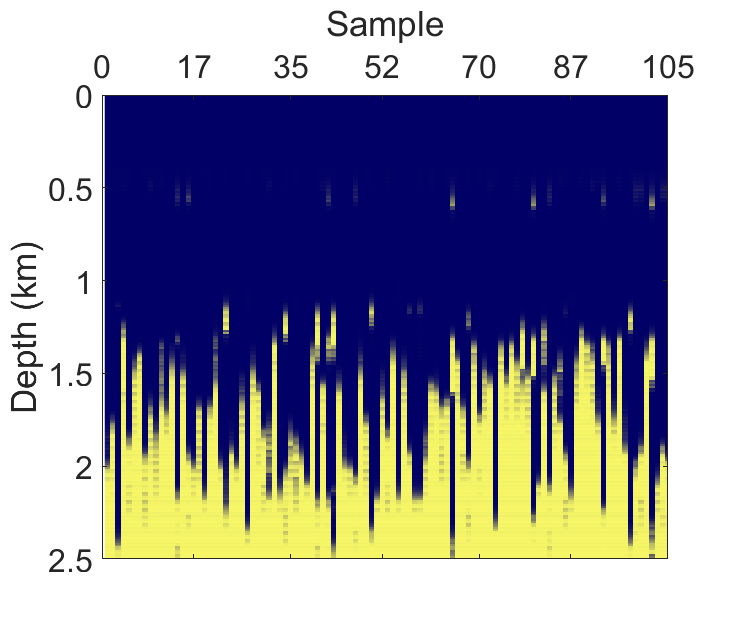}
\end{minipage}}
\subfigure[SEM: Difference]
{\begin{minipage}[t]{0.33\linewidth}
\centering
\includegraphics*[scale = .25]{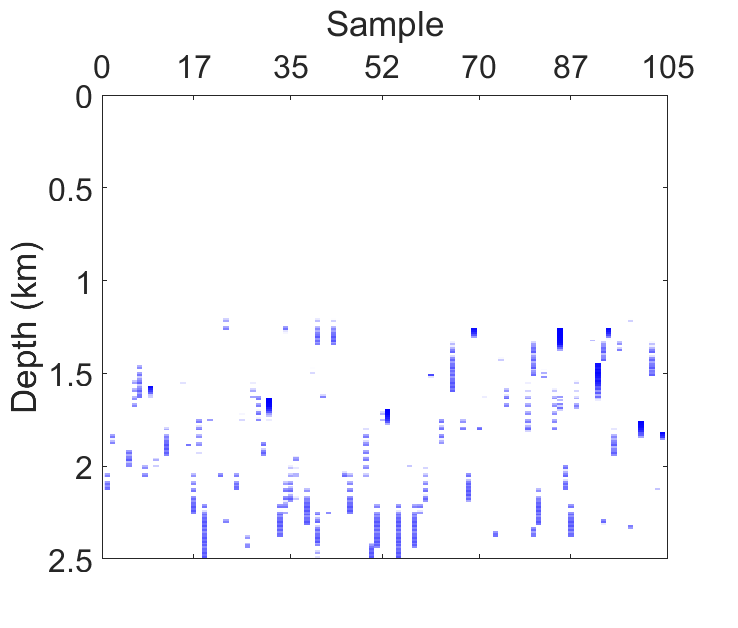}
\end{minipage}}
\caption{1D interface predicted by GeoDUDe-3 using P-wave data. Each column of pixels represents a data point. The value of each pixel describes whether the material at the depth corresponding to that pixel's column belongs to either the high or low velocity region. The blue pixels represent the low velocity region, while the yellow represent the high velocity region. Subfigures (c), (f) show the difference between the predicted and actual velocity profile, where the accuracy is measured by the wrong-labeled pixels (blue) over the total number of pixels in the figures (c), (f). In fact, after 3500 epochs of training GeoDUDe-3 achieved a 96.97\% evaluation accuracy on data generated by the FGA.}

\label{fig:p_results}
%\vspace{-10pt}
\end{figure}

\begin{figure} 
\subfigure[FGA]
{\begin{minipage}[t]{0.5\linewidth}
\centering
\includegraphics*[scale = .35]{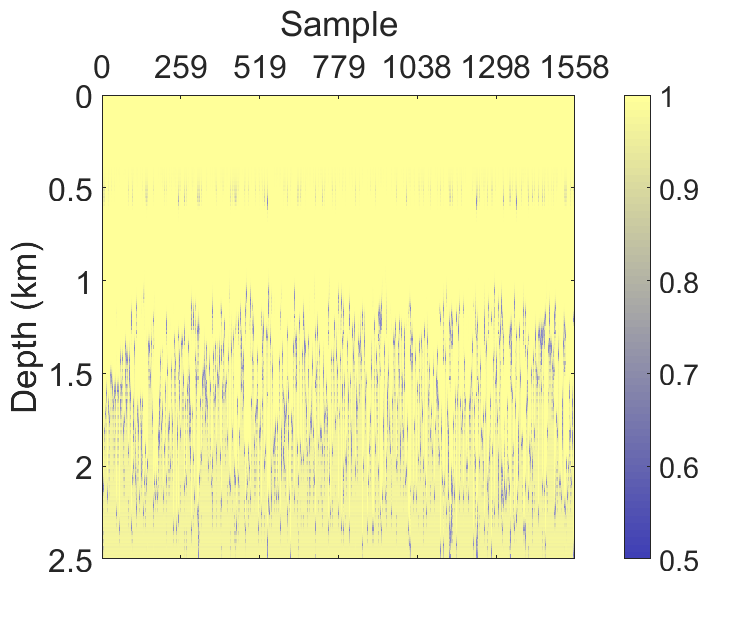}
\end{minipage}}
\subfigure[SEM]
{\begin{minipage}[t]{0.5\linewidth}
\centering
\includegraphics*[scale = .35]{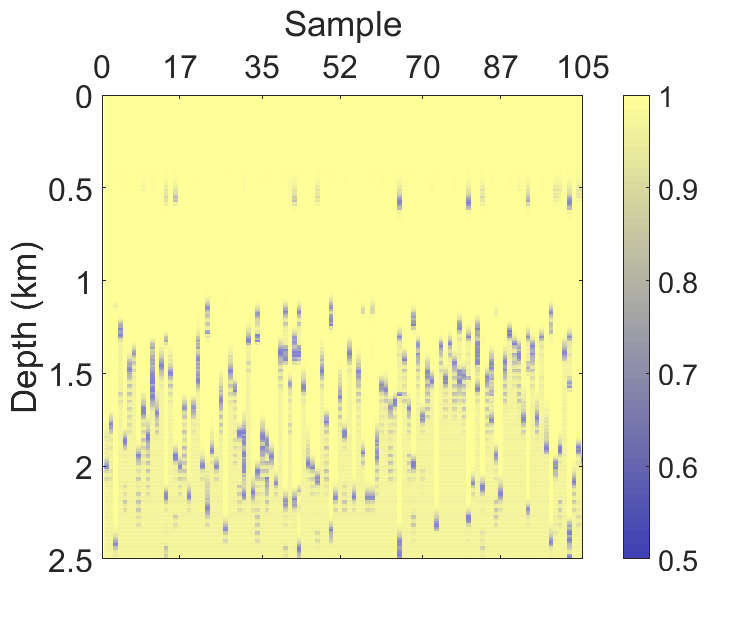}
\end{minipage}}
\caption{P-wave confidence distribution comparison produced by GeoDUDe-3 for 1D interface problem. Regions of low confidence correspond to areas where an interface is likely. The color bar is a probability spectrum from 0.5 to 1.  In general, the closer the network gets to the interface the less confident its prediction becomes.}
%Recall the heatmap is a sample is the vector of probabilities that a depth indicated by the index belongs is correctly classified.}
\label{fig:p_heat}
\end{figure}

After 3500 epochs of training GeoDUDe-3 achieved a 96.97\% evaluation accuracy on data generated by the FGA. When evaluated on data generated by SPECFEM dataset GeoDUDe-3 achieved a 94.29\% evaluation accuracy, only a 2.68\% decrease. We remark in \cite{AO_Fawzi_Frossard_UniAdPert,Szegedy_Zaremba_Sutskever_intriguingNN}, it was shown even small perturbations in input can affect network classification results. This suggests that the asymptotic errors present in the FGA do not greatly affect the segmentation problem. Visualizations of the output for GeoDUDe-3 are shown in Figure~\ref{fig:p_results}. Figure~\ref{fig:p_heat} shows the heatmap. Recall this displays the confidence the network places on the pixels prediction.

\subsubsection{P,S-wave data set}

\begin{figure} 
\subfigure[Computational Domain]
{\begin{minipage}[t]{0.5\linewidth}
\centering
\includegraphics*[scale = .35]{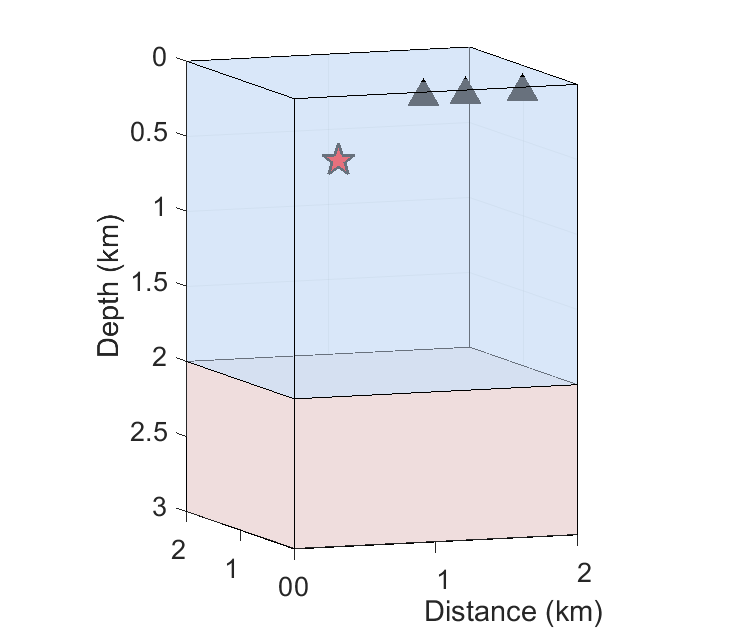}
\end{minipage}}%%
\subfigure[Seismogram]
{\begin{minipage}[t]{0.5\linewidth}
\centering
\includegraphics*[scale = .35]{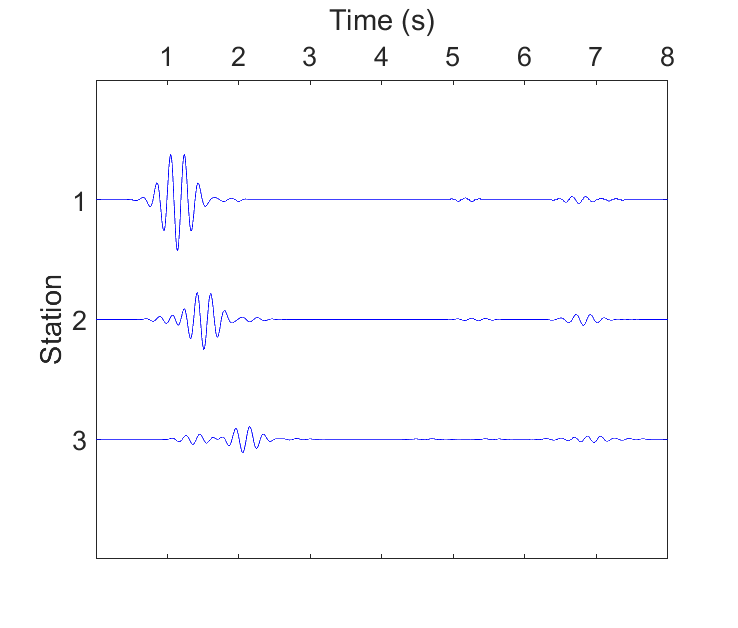}
\end{minipage}}
\caption{The locations of source and receivers, and the generated synthetic P- and S-wave seismograms for the 1D interface problem. We take $k=32$ for generating the synthetic data. (a) The source is located at (.5,.5,.5) as a star and the 3 receivers are located on the surface. The interface presented is at a depth of 2 km. (b) A visualization of typical data point, which is a collection of 3 seismograms from the forward simulation using the FGA.}\label{fig:ps:setup}
\end{figure}

\textbf{Dataset.} The P,S-wave dataset is generated with a computation domain of $[0,2]$ km$\times[0,2]$ km$\times[0,3]$ km with a source centered at $\bx_0 = (0.5, 0.5, 0.5)$ km, and wavenumber $k = 32$, or approximately $5.09$ Hz. The stations lie in a plane and are located just below the surface at $S_1:(1.1,0.5,0.1)$ km, $S_2:(1.4,0.5,.1)$ km, $S_3:(1.8,0.5,0.1)$ km.  The interface is a plane, $z = z_0$ that varies from depth 1 km to 2 km. Above the interface $c_\tp$ varies linearly from 0.75 km/s to 1.10 km/s, below the interface $c_\tp$ varies linearly from 1.12 km/s to 1.48 km/s and we fix $c_\ts = c_\tp/1.7$ (corresponding the case $\lambda \approx \mu$). See Figure \ref{fig:ps:setup}. There are a total of 6,400 data points in the P,S-wave dataset. Each data point is a (2048,3,3) tensor. Prior to training each example is down-sampled along the temporal axis by a factor of 8. Each network used a mini-batch training size of 256. Similarly to the P-wave dataset, 100 additional samples were generated using SPECFEM3D for evaluation after training. \\

{\bf Network Details. } GeoDUDe-2 and GeoDUDe-3 with default parameters were used. Both networks were trained using a NADAM optimizer with a dropout probability of 0.5. \\

\begin{figure} 
\subfigure[Evaluation Accuracy ]
{\begin{minipage}[t]{0.5\linewidth}
\centering
\includegraphics*[scale = .35]{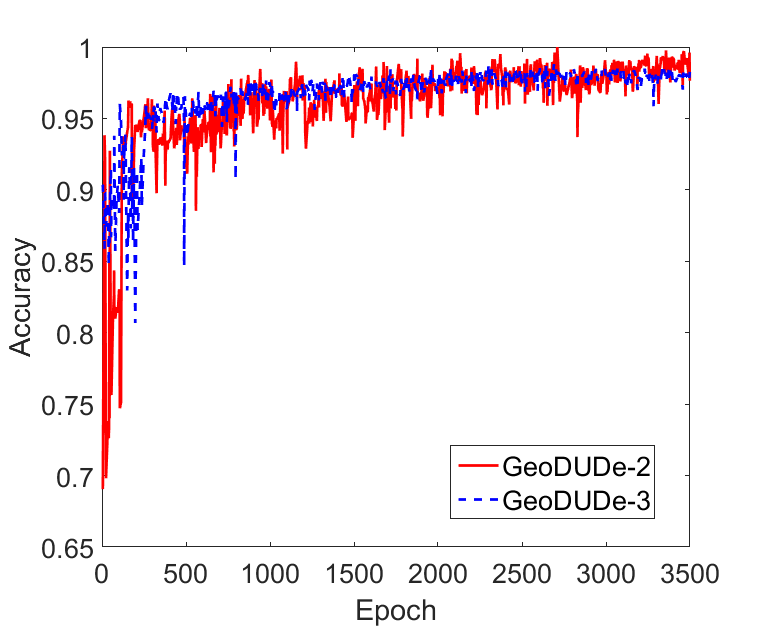}
\end{minipage}}%%
\subfigure[Training Accuracy]
{\begin{minipage}[t]{0.5\linewidth}
\centering
\includegraphics*[scale = .35]{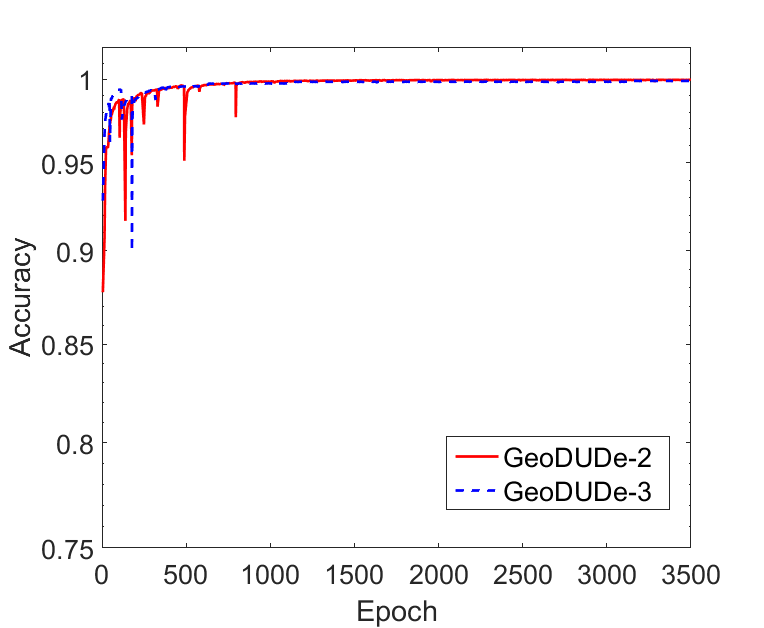}
\end{minipage}}
\caption{PS-wave training results for 1D interface problem, with synthetic data generated for $k=32$ in \eqref{eq:srcfun}: The evaluation data set for this figure only contains data generated by the FGA. }\label{fig:ps:eval}
\end{figure}

\begin{figure} 
  \subfigure[FGA: Actual]
  {\begin{minipage}[t]{0.33\linewidth}
  \centering
  \includegraphics*[scale = .25]{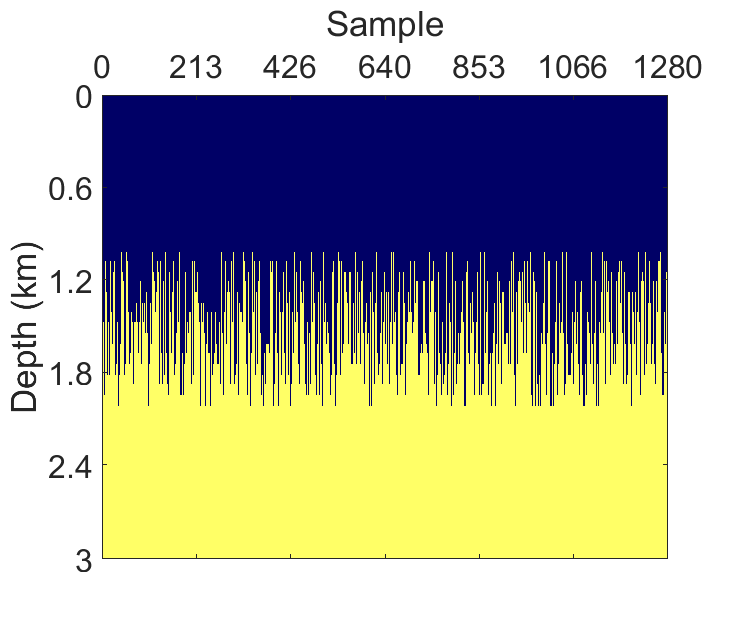}
  \end{minipage}}%%
  \subfigure[FGA: Predicted]
  {\begin{minipage}[t]{0.33\linewidth}
  \centering
  \includegraphics*[scale = .25]{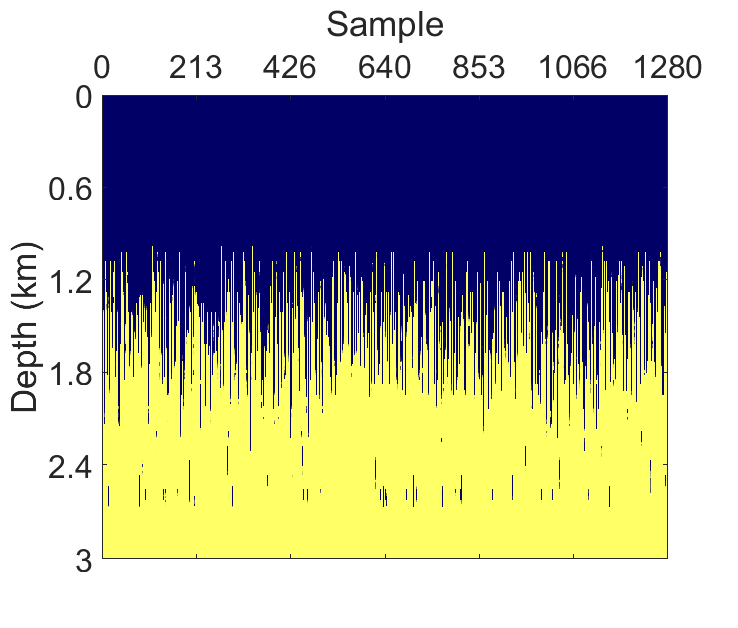}
  \end{minipage}}
  \subfigure[FGA: Difference]
  {\begin{minipage}[t]{0.33\linewidth}
  \centering
  \includegraphics*[scale = .25]{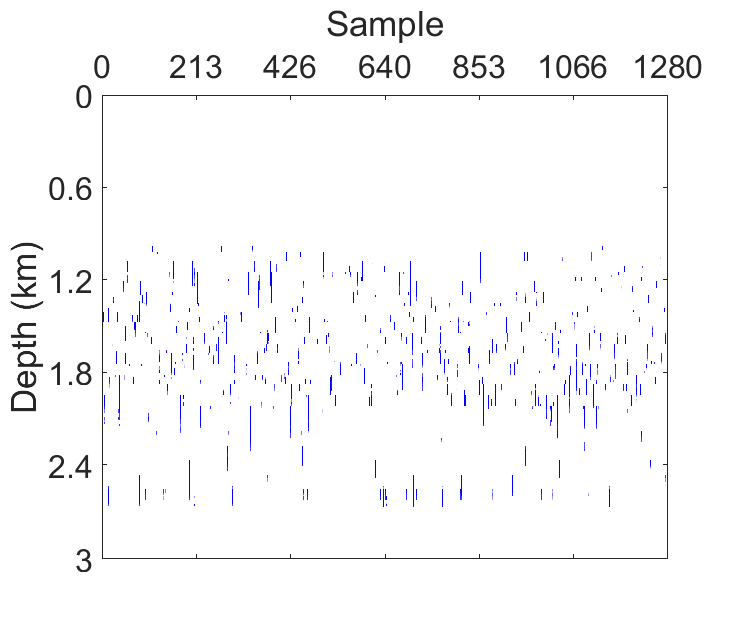}
  \end{minipage}}%%
	\\
	\subfigure[SEM: Actual]
	{\begin{minipage}[t]{0.33\linewidth}
\centering
\includegraphics*[scale = .25]{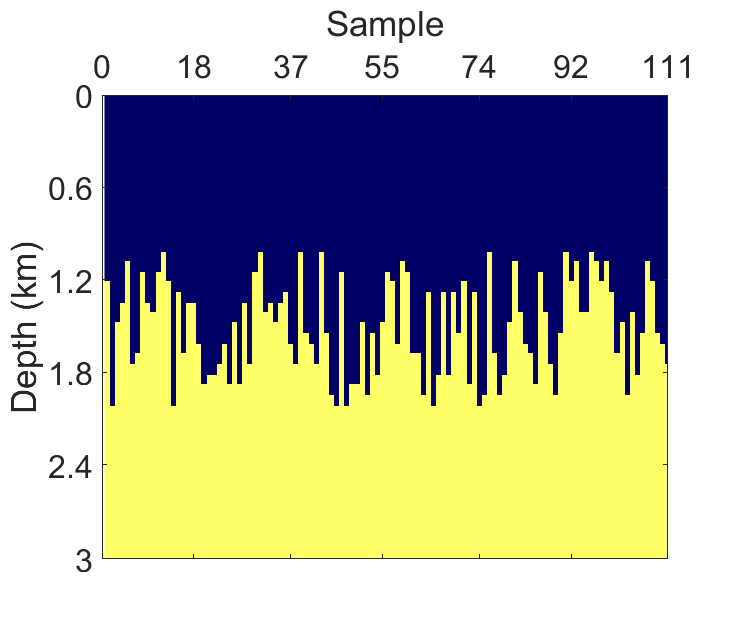}
\end{minipage}}%%
\subfigure[SEM: Predicted]
{\begin{minipage}[t]{0.33\linewidth}
\centering
\includegraphics*[scale = .25]{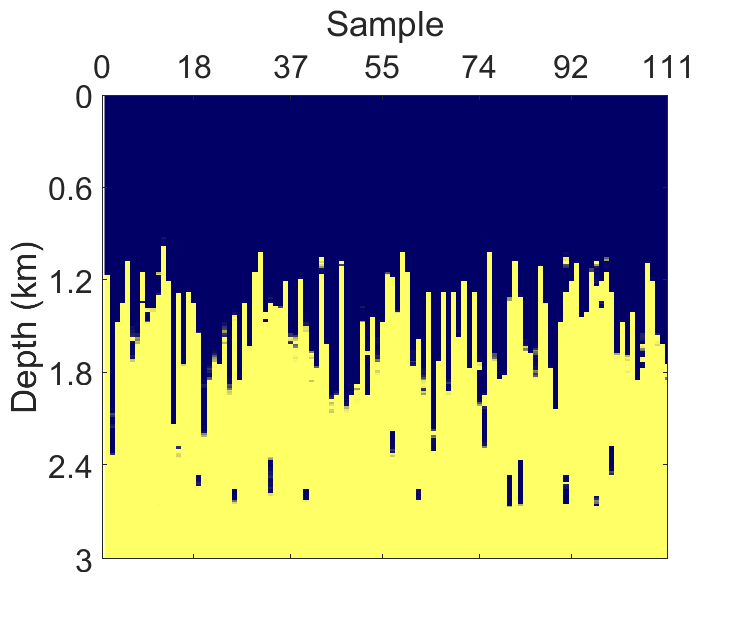}
\end{minipage}}
\subfigure[SEM: Difference]
{\begin{minipage}[t]{0.33\linewidth}
\centering
\includegraphics*[scale = .25]{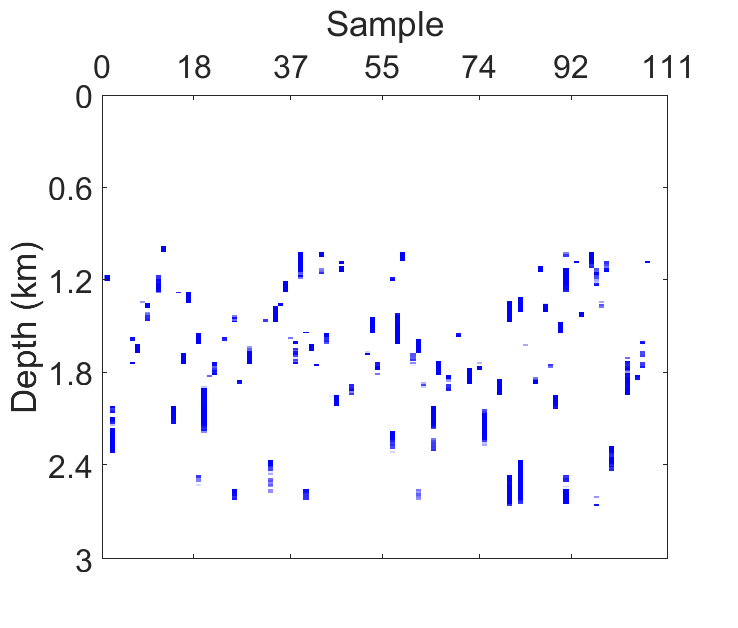}
\end{minipage}}%%
\caption{1D interface predicted by GeoDUDe-2 using P,S-wave data. Each column of pixels represents a sample. The value of each pixel describes whether the material at the depth corresponding to that pixel's column belongs to either the high or low velocity region. The blue pixels represent the low velocity region, while the yellow represent the high velocity region. Subfigures (c), (f) show the difference between the predicted and actual velocity profile, where the accuracy is measured by the wrong-labeled pixels (blue) over the total number of pixels in the figures (c), (f). We give the statistical accuracy in Table~\ref{tab:PS-Network-Comparison}, which shows an accuracy of over $96\%$.}\label{fig:ps_results}
\end{figure}

\begin{figure} \label{fig:13}
\subfigure[GeoDUDe-2:FGA]
{\begin{minipage}[t]{0.50\linewidth}
\centering
\includegraphics*[scale = .35]{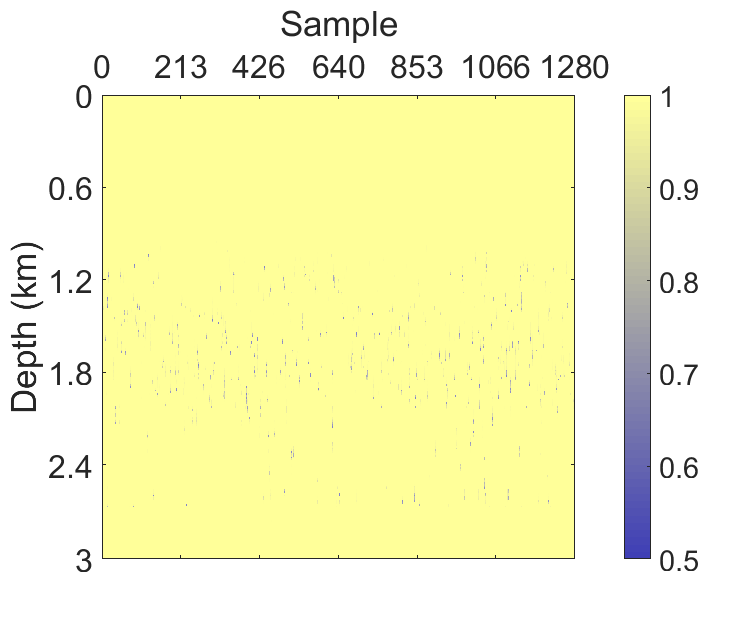}
\end{minipage}}
\subfigure[GeoDUDe-2:SEM]
{\begin{minipage}[t]{0.50\linewidth}
\centering
\includegraphics*[scale = .35]{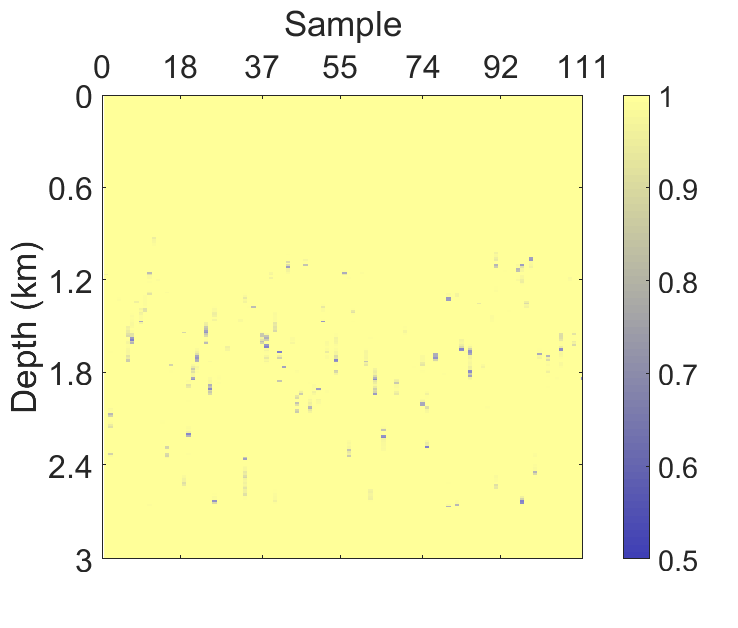}
\end{minipage}}
\caption{P,S-wave heat-map distribution comparison produced by GeoDUDe-2 for 1D interface problem. Regions of low confidence correspond to areas where an interface is likely. The color bar is a probability spectrum from 0.5 to 1. In general, the closer the network gets to the interface the less confident its prediction becomes.
}\label{fig:ps_heat}
\end{figure}

{\bf Results. } Both networks were trained for 3500 epochs. The most successful network was GeoDUDe-2, with 98.26 \% evaluation accuracy on FGA data, and 97.55 \% evaluation accuracy on the SPECFEM data . We find that the evaluation accuracy goes down for deeper networks. In particular, GeoDUDe-3 performed worse with only a 92.34 \% evaluation accuracy, especially compared to the same network architecture on the P-wave dataset. This is likely due to overfitting of the data causing an increase in generalization error. Similarly to the P-wave dataset, evaluation accuracies on SPECFEM3D data are only marginally worse than their FGA counterparts, with a max difference of 1.17\% between the datasets. See Table \ref{tab:PS-Network-Comparison} for the summary of the results and Figures~\ref{fig:ps:eval}, \ref{fig:ps_results} and \ref{fig:ps_heat}.

%\begin{table}
	%\begin{center}
		%\caption{P,S-Data Network Comparisons.}
		%\label{tab:PS-Network-Comparison}
		%\begin{tabular}{r|r|r|r|r|r}
			%\textbf{Network} & \textbf{Eval. Acc.} & %\textbf{Eval. Loss} & \textbf{Train. Acc.} & %\textbf{Train. Loss} & \textbf{SEM Acc.} \\
			%\hline
			%GeoDUDe-2 & 98.26 \% & 0.1650 & 99.97 \% & 0.0015729 %& 97.55 \% \\
			%\hline
			%GeoDUDe-3 & 97.64 \% & 0.1975 & 99.90 \% & 0.0026064 %& 96.47 \% \\
	%	\end{tabular}
	%\end{center}
%\end{table}

\begin{table}
	\begin{center}
		\caption{P,S-Data Network Comparisons for 1D interface problem. Here Eval. Acc. $=$ evaluation accuracy, Train. Acc. $=$ training accuracy, and SEM Acc. $=$ evaluation accuracy tested by SEM synthetic data.}
		\label{tab:PS-Network-Comparison}
		\begin{tabular}{r|r|r|r}
			\textbf{Network} & \textbf{Eval. Acc.}  & \textbf{Train. Acc.}  & \textbf{SEM Acc.} \\
			\hline
			GeoDUDe-2 & 98.26 \% & 99.97 \% & 97.55 \% \\
			\hline
			GeoDUDe-3 & 97.64 \%  & 99.90 \%  & 96.47 \% \\
		\end{tabular}
	\end{center}
\end{table}

%%%%%%%%%%%%%%%%%%%%%%%%%%%
%\section{Experiments and Results} \label{E&R}
%%%%%%%%%%%%%%%%%%%%%%%%%%%%
%In this section we use GeoSeg to address three more complicated models: a three layer media, a three layer media with a low velocity pocket in the middle layer, and a three layer media with a low velocity region formed as the union of possibly overlapping cylindrical pockets. For all following experiments 32 colinear receivers are located on the surface with coordinates  $S_i = (2(i - 1)/31, 1, 0)$ km. The source and wave speeds will vary for each test and the frequency will be fixed at $k = 64$. The data for these tests is generated on the clusters \emph{knot} and \emph{pod}, at the center for scientific computing at UC Santa Barbara. The training was performed on a single NVIDIA V100 GPU on the \emph{pod} cluster.

\subsection{Three-Layered Media}

\begin{figure} \label{fig:14}
\subfigure[Computational domain]
{\begin{minipage}[t]{0.50\linewidth}
\centering
\includegraphics*[scale = .35]{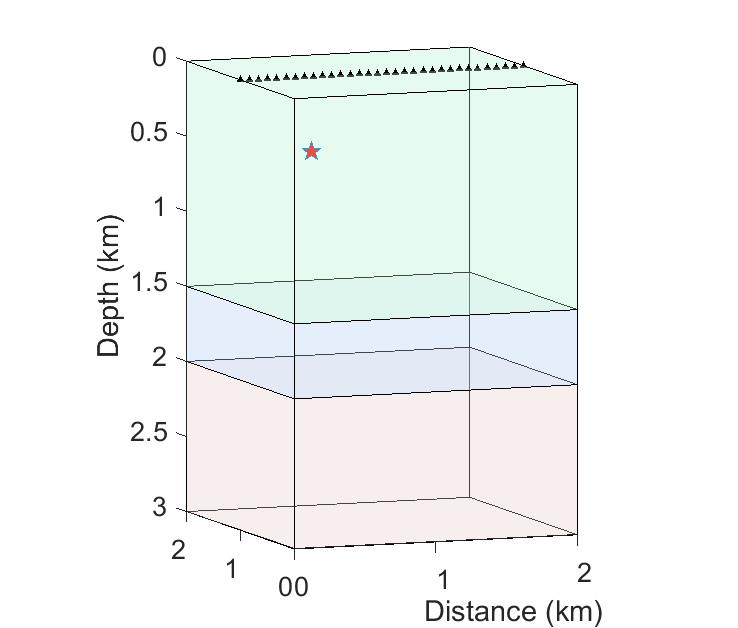}
\end{minipage}}
\subfigure[seismograph]
{\begin{minipage}[t]{0.50\linewidth}
\centering
\includegraphics*[scale = .35]{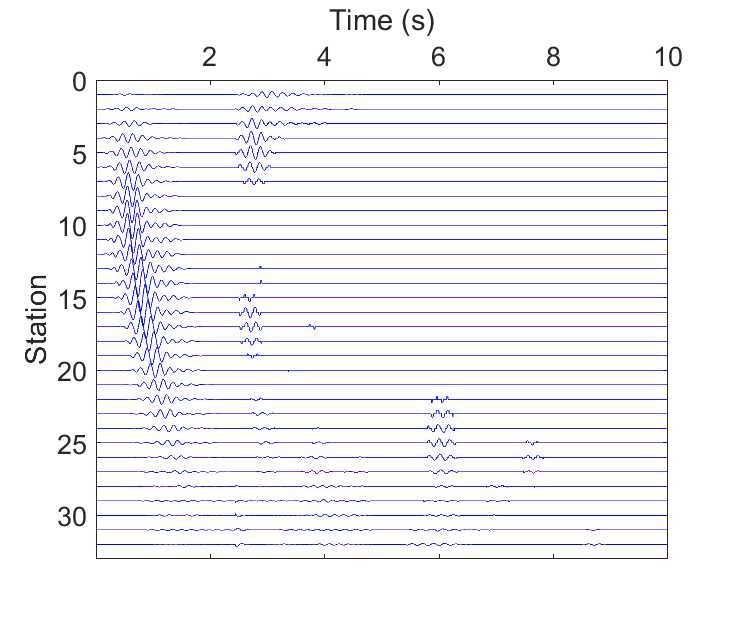}
\end{minipage}}
\caption{The locations of source and receivers, and the generated synthetic P- and S-wave seismograms for the three-layered media model. We take $k=32$ for generating the synthetic data. (a) The source is located at $(.5,1,.5)$ km as a star, the 32 receivers are located on the surface on the plane $y = 1$ km, and the interfaces presented are at a depth of 1.5 km and 2 km. (b) A visualization of typical data point, which is a collection of 32 seismograms from the forward simulation using the FGA.}\label{fig:3L_setup}
\end{figure}

{\bf Dataset. } A natural extension of the model is to include one or more low velocity regions in the computational domain. For this experiment we consider a three-layered media with a low velocity region in the middle, the velocities in each region will be fixed.The P-wave speed is  $c_{\tp} = 1.3, 0.9, 1.7$ km/s for the top, middle, and bottom layers, receptively.  The S-wave speed is set to $c_{\ts}= c_{\tp}/1.7$ for each layer. The lower interface will be in a rage of 1.8 km and 2.8 km by an increment of 1 m. Similarly the upper interface will vary from .2 km to 1.2 km by an increment of 1m. See Figure \ref{fig:3L_setup}. There were 10201 samples with a batch size of 64. \\

{\bf Network Details. } GeoDUDe-3 was used. During training the dropout probability was 0.12. Training was performed with stochastic gradient descent with a learning rate of 0.001. \\

\begin{figure} 
  \subfigure[Actual]
  {\begin{minipage}[t]{0.33\linewidth}
  \centering
  \includegraphics*[scale = .25]{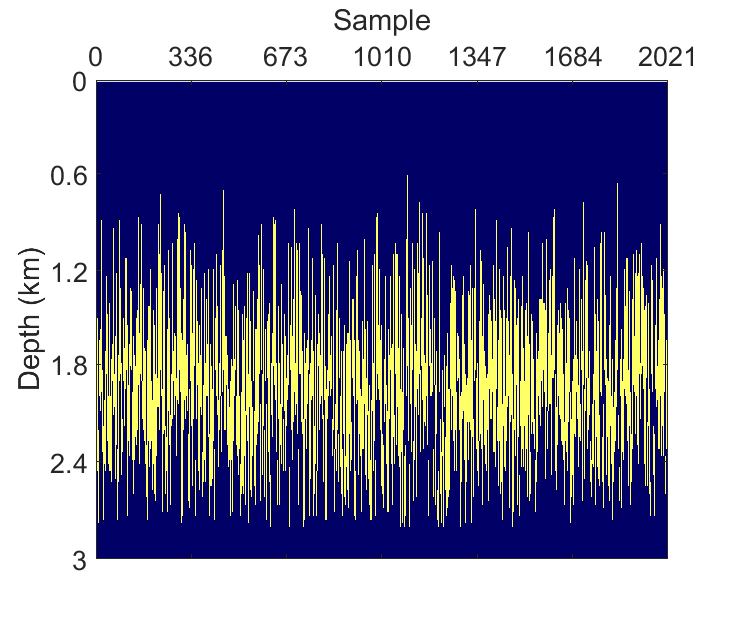}
  \end{minipage}}%%
  \subfigure[Predicted]
  {\begin{minipage}[t]{0.33\linewidth}
  \centering
  \includegraphics*[scale = .25]{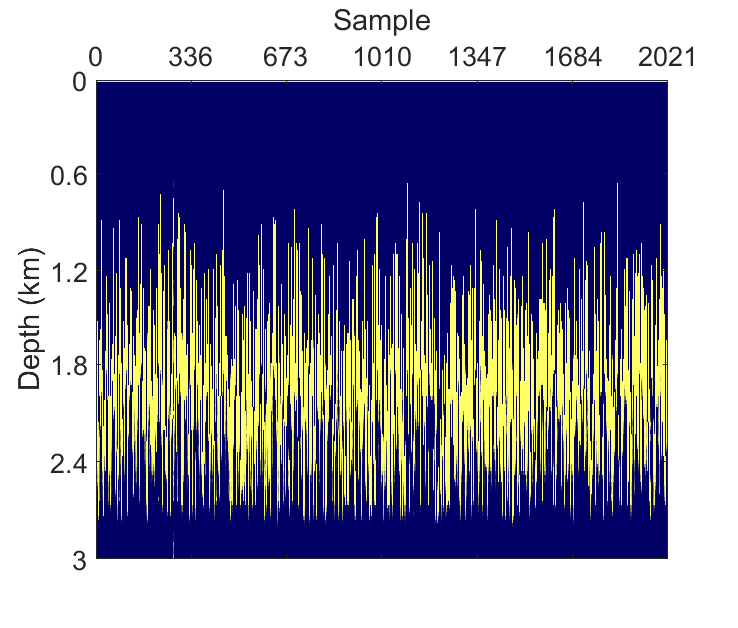}
  \end{minipage}}
  \subfigure[Difference]
  {\begin{minipage}[t]{0.33\linewidth}
  \centering
  \includegraphics*[scale = .25]{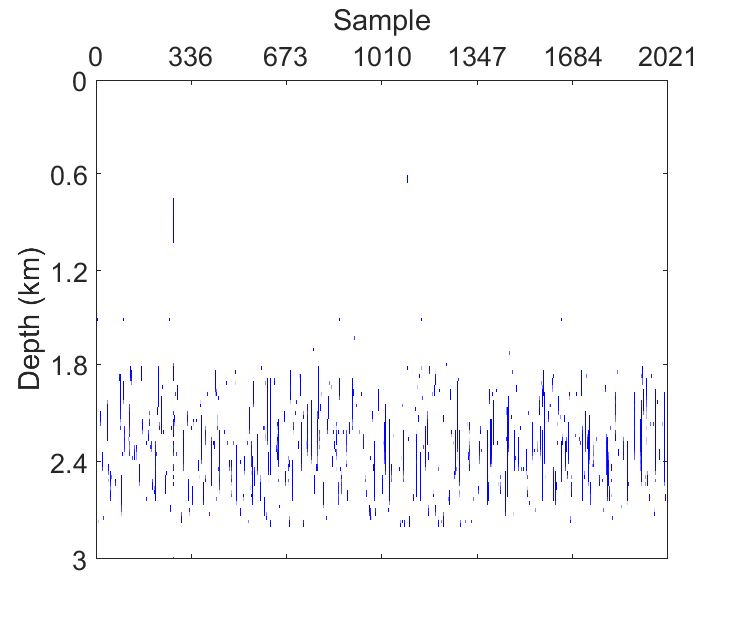}
  \end{minipage}}%%
\caption{Predictions for three-layered media by GeoDUDe-3: Each column of pixels represents a sample. The value of each pixel describes whether the material at the depth corresponding to that pixel's column belongs to either the high or low velocity region. The color bar is a probability spectrum from 0.5 to 1.  In general, the closer the network gets to the interface the less confident its prediction becomes.  There is a slight loss of confidence for the network detecting the lower interface.
}\label{fig:3L_results}
\end{figure}

\begin{figure}
  \centering
  \includegraphics*[scale = .65]{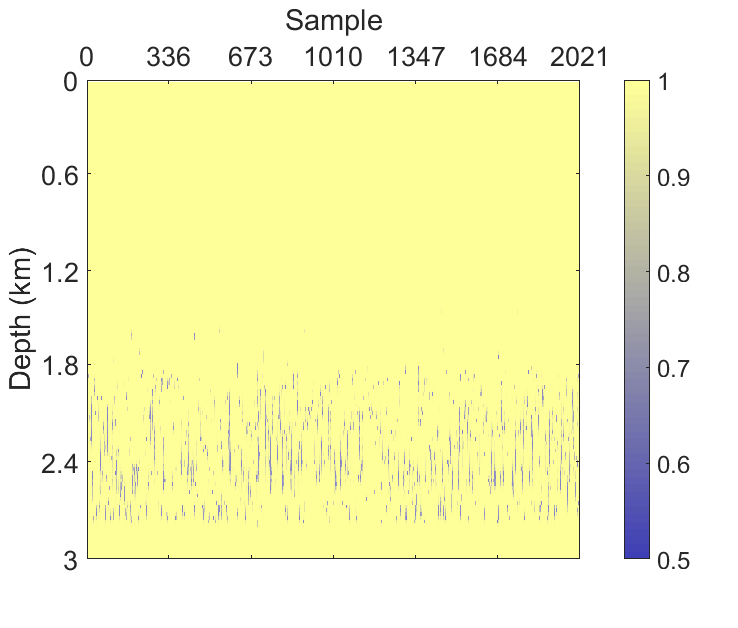}
\caption{Confidence map for three-layered media model produced by GeoDUDe-3. Regions of low confidence correspond to areas where an interface is likely. The color bar is a probability spectrum from 0.5 to 1.  In general, the closer the network gets to the interface the less confident its prediction becomes.
}\label{fig:3L_heat}
\end{figure}

{\bf Results. } The network achieved a training accuracy of 99.51\% and an evaluation accuracy of 95.51\% after 3000 epochs. See Figures \ref{fig:3L_results} and \ref{fig:3L_heat}.


%%%%%%%%%%%%%%%%%%%%
%%%%%%%%%%%%%%%%%%%%
%%%%%%%%%%%%%%%%%%%%

\subsection{Sine interface model}

\begin{figure} \label{fig:sine_1}
\subfigure[Computational domain]
{\begin{minipage}[t]{0.50\linewidth}
\centering
\includegraphics*[scale = .35]{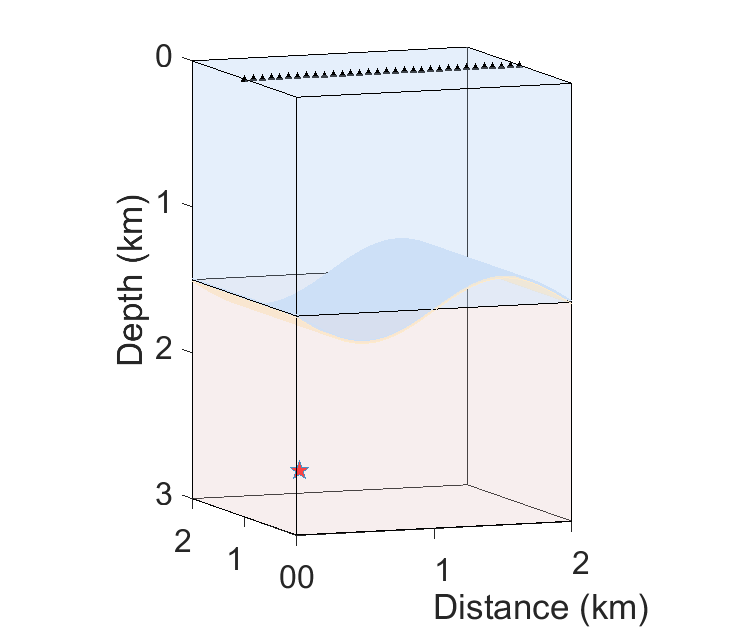}
\end{minipage}}
\subfigure[Displacement color channel]
{\begin{minipage}[t]{0.50\linewidth}
\centering
\includegraphics*[scale = .35]{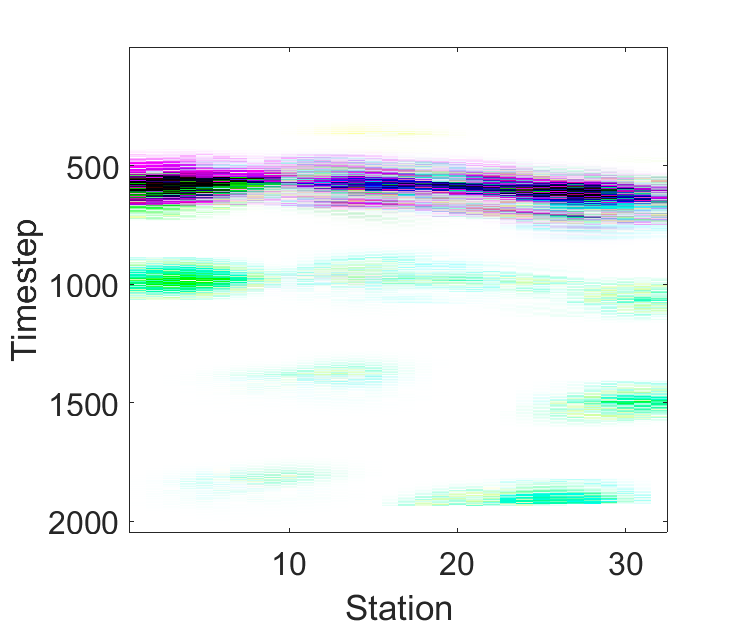}
\end{minipage}}
\caption{The locations of source and receivers, and the generated synthetic P- and S-wave seismograms for the sine interface model. We take $k=64$ for generating the synthetic data. (a) The source is located at $(0.4, 1, 2.7)$ km as a star, the 32 receivers are located on the surface on the plane $y = 1$ km, and the interfaces presented are at a depth of 1.5 km. (b) Visualization of network input as image for sine interface model. Each color channel (inverse RGB) represents a coordinate of the displacement.}\label{fig:sine_setup}
\end{figure}

{\bf Dataset. }  For this experiment we consider a more complicated interface, which is given by the level set function; $F(x,z) = z - 0.1\sin(\pi f x)$. The level set values $F(x,z) = D$ vary between 0.3 km and 2.3 km by an increment of $.001$ km.  And the phase factor $f$ will ranges from 1 to 2 by an increment of $.001$. The nondimensionalized is set to wavenumber $k = 64$ and $c_\tp = 1.1, 1.5$ for the top and bottom layers will all be fixed. As before, $c_\ts$ will be a fixed multiple of $c_\tp$ by 1.7.  We record the displacement for $10$ s; see Figure~\ref{fig:sine_setup} for source, receiver details.

{\bf Network Details. } GeoDUDe-4 was used with a dropout probability of 0.2. Training was performed with 1500 epochs using the NADAM optimizer followed by 1000 epochs of stochastic gradient descent with a learning rate of 0.001.  There were 10050 date points generated.  9150 data points are used for training with a mini-batch size of 25; 900 data points were used for evaluation. 

\begin{figure} 
  \subfigure[Actual]
  {\begin{minipage}[t]{0.33\linewidth}
  \centering
  \includegraphics*[scale = .25]{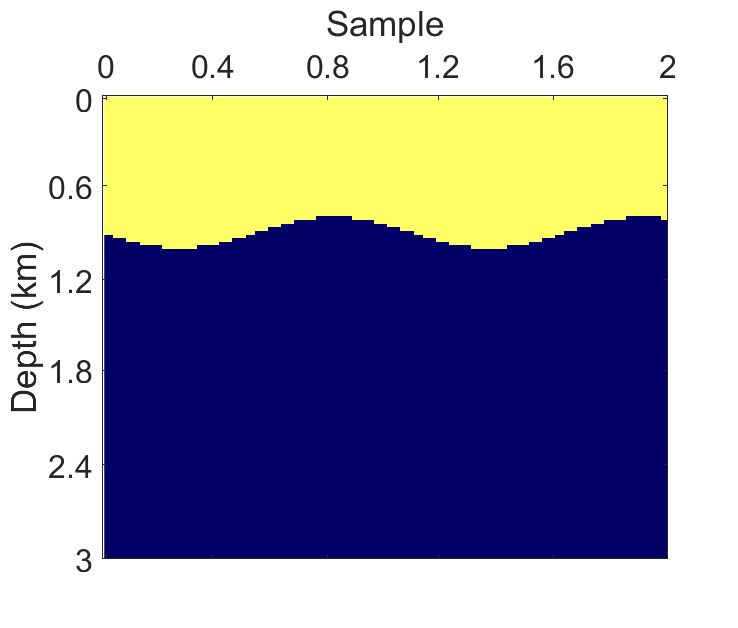}
  \end{minipage}}%%
  \subfigure[Predicted]
  {\begin{minipage}[t]{0.33\linewidth}
  \centering
  \includegraphics*[scale = .25]{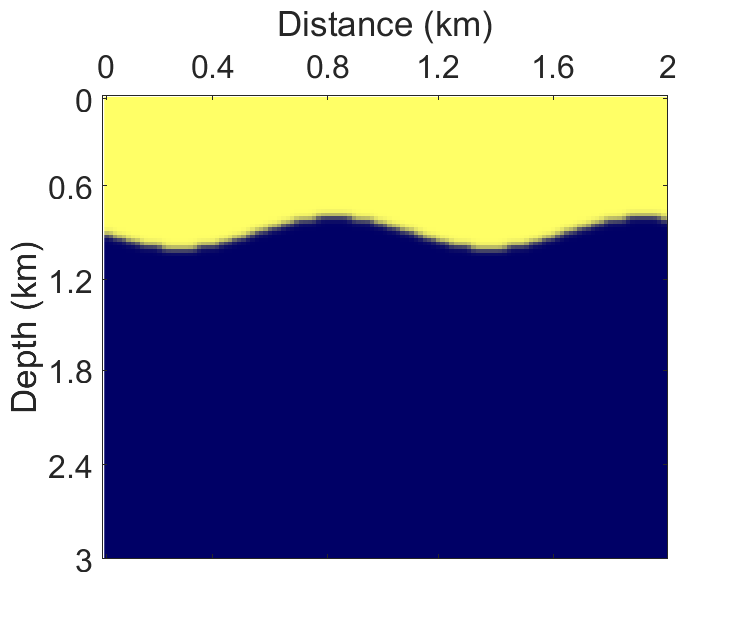}
  \end{minipage}}
  \subfigure[Confidence map]
  {\begin{minipage}[t]{0.33\linewidth}
  \centering
  \includegraphics*[scale = .25]{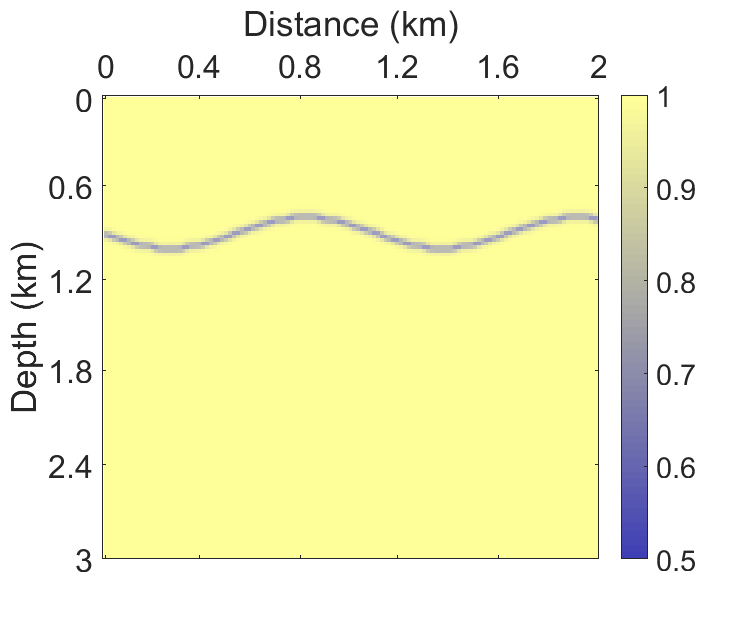}
  \end{minipage}}%%
\caption{Typical results from training phase factor $f =  1.83$,  level set value $D = .8898$. (a), (b) Ground truth and prediction for sine interface GeoDUDe-4. (c) Confidence map for sine interface model. Regions of low confidence correspond to areas where an interface is likely.}\label{fig:sine_results}
\end{figure}

{\bf Results. } The network achieved a training accuracy of 99.92\% and an evaluation accuracy of 99.28\% after 2500 epochs. See Figure \ref{fig:sine_results} for evaluation of a typical data point.


%%%%%%%%%%%%%%%%%%%%
%%%%%%%%%%%%%%%%%%%%
%%%%%%%%%%%%%%%%%%%%

\subsection{2D Low Velocity Pocket}\label{2dp:section}

\begin{figure} 
  \subfigure[Computational domain]
  {\begin{minipage}[t]{0.5\linewidth}
  \centering
  \includegraphics*[scale = .35]{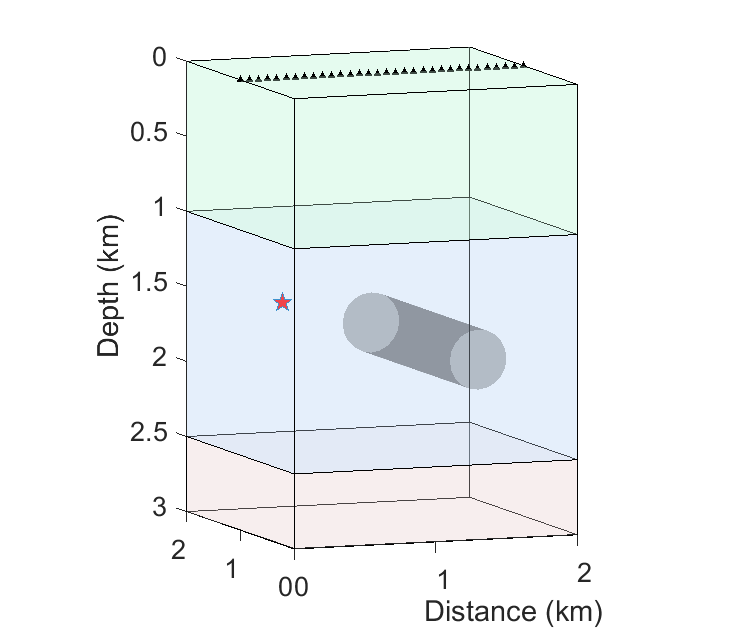}
  \end{minipage}}%%
  \subfigure[Displacement color channel]
  {\begin{minipage}[t]{0.5\linewidth}
  \centering
	\includegraphics*[scale = .35]{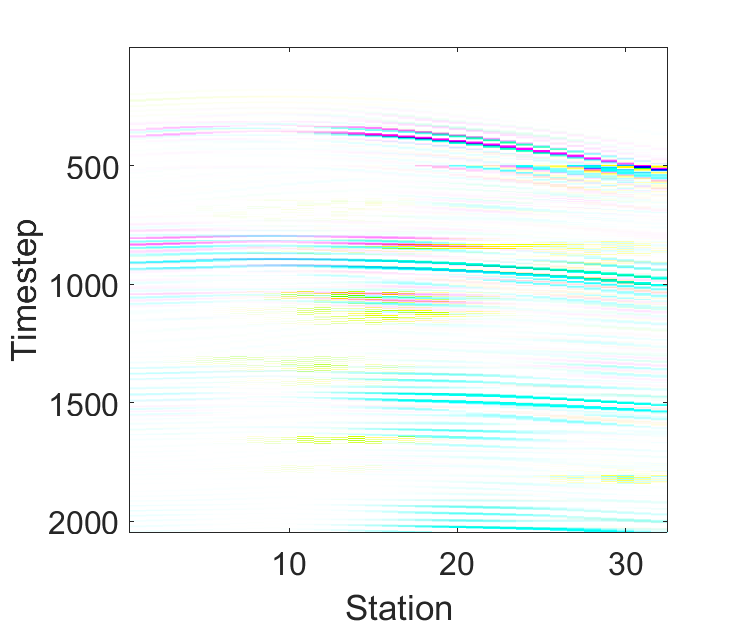}
  \end{minipage}}
\caption{The locations of source and receivers, and generated synthetic P- and S-wave seismograms for the 2D pocket model. We take $k=32$ for generating the synthetic data. (a) The  source is located at $(.5,1,1.5)$ km as a star and the receivers are located on the surface on the plane $y = 1$ km. The interfaces are fixed at a depth of 1 km and 2.5 km. Visualization of network input as image for 2D pocket model. Each color channel (inverse RGB) represents a coordinate of the displacement.}\label{fig:2dp_color}%(b) A visualization of typical data point, which is a collection of 32 seismograms from the forward simulation using the FGA.}
\label{fig:2dp_cd_seis}
\end{figure}

\textbf{Dataset:} We now investigate whether the network can learn more complex 2D geometries.  The considered models each will be a three-layered problem with a low velocity cylindrical region in the middle layer. The source will be located at $(.5,1,1.5)$ km.  The interfaces located at 1 km and 2.5 km will be fixed. A cylinder with center $(x,z)$ and radius $r$ will be randomly generated $x\in [0.85 ,1.65]$ km,  $z\in [1.35, 2.15]$ km, and $r \in [.05,.3]$ km with samples taken from a uniform distribution. See Figure~\ref{fig:2dp_cd_seis}. 11350 data points are generated with 1000 being saved for evaluation. The P-wave speeds will be fixed and are $c_{\tp} = 1.1, 1.3, 1.7$ km/s, for the top, middle and bottom layers respectively. The S-wave speed, $c_{\ts}$ will be a fixed multiple of $c_{\tp}$ by 1.7 for each layer.  Inside the pocket the P-wave speed is set to $c_{\tp} = 0.5$ km/s and the S-wave speed is set to zero, $c_{\ts} = 0$.  Only $P$-waves will propagate through the cylinder; However, S-wave can transmit to P-wave going in the pocket and P-wave can transmit to P,S-waves coming out of the pocket. Unlike previous models the goal is to identify a low velocity region in a three layered media in a 2D slice of the computational domain. A batch size of 20 examples was used.  \\

{\bf Network Design. } A GeoDUDe-4 network was used with a two layer transfer branch before its input. The dropout probability was 0.2. \\

\begin{figure} 
  \subfigure[x-coordinate]
  {\begin{minipage}[t]{0.33\linewidth}
  \centering
  \includegraphics*[scale = .25]{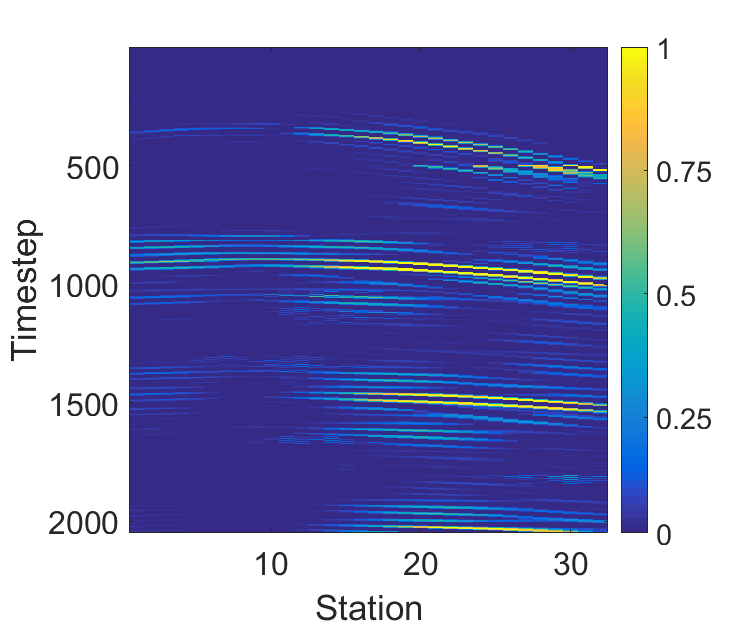}
  \end{minipage}}%%
  \subfigure[y-coordinate]
  {\begin{minipage}[t]{0.33\linewidth}
  \centering
  \includegraphics*[scale = .25]{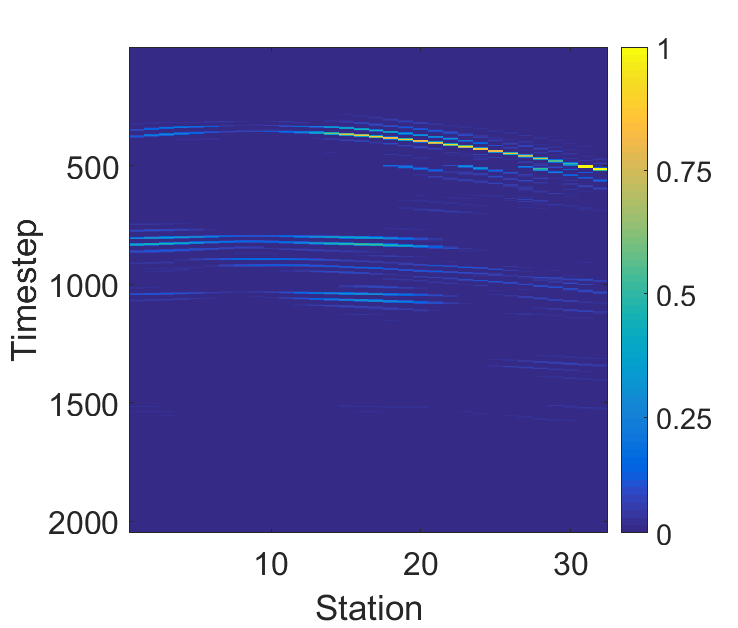}
  \end{minipage}}%%
  \subfigure[z-coordinate]
  {\begin{minipage}[t]{0.33\linewidth}
  \centering
  \includegraphics*[scale = .25]{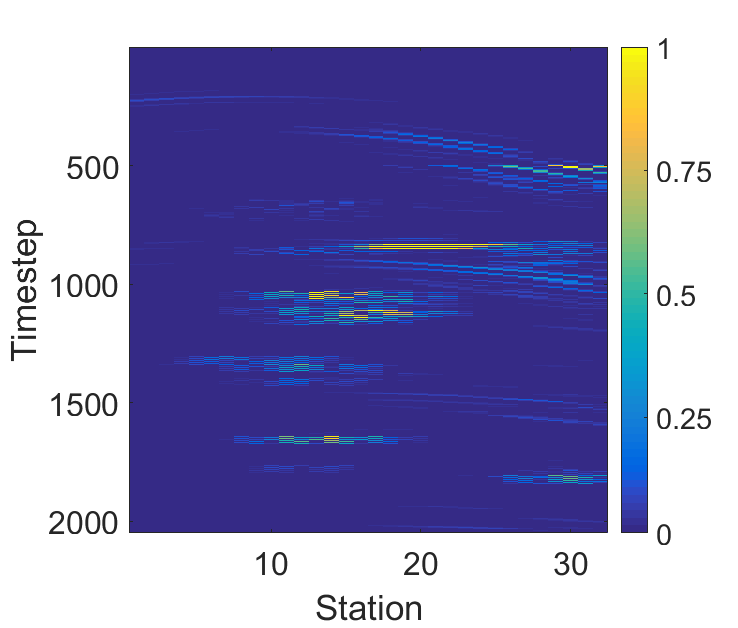}
  \end{minipage}}
\caption{Visualization of network input using normalized displacement data for 2D pocket model.}\label{fig:2dp_disp}
\end{figure}

\begin{figure}
  \subfigure[Actual]
  {\begin{minipage}[t]{0.33\linewidth}
  \centering
  \includegraphics*[scale = .25]{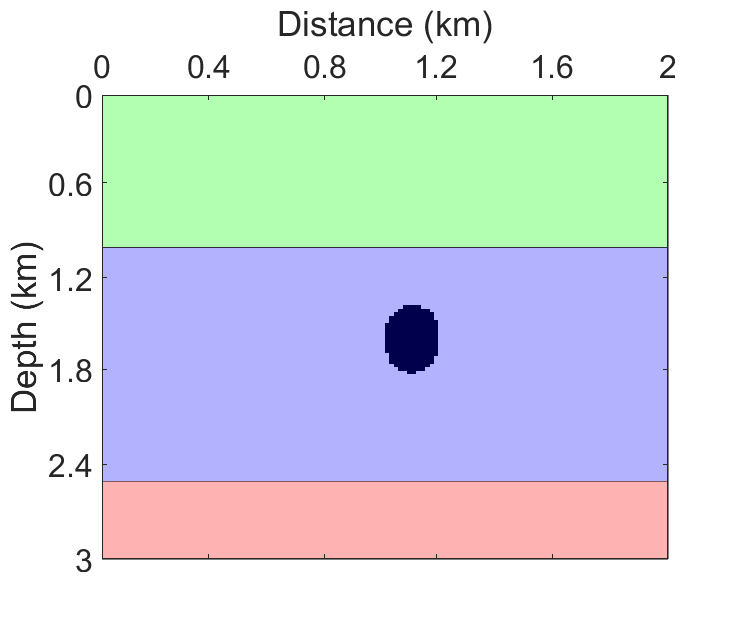}
  \end{minipage}}%%
  \subfigure[Predicted]
  {\begin{minipage}[t]{0.33\linewidth}
  \centering
  \includegraphics*[scale = .25]{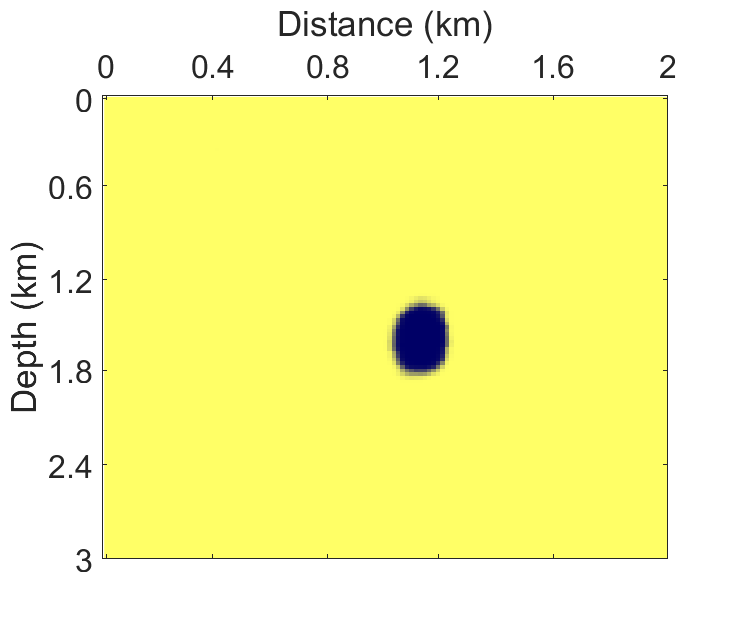}
  \end{minipage}}%%
  \subfigure[Confidence map]
  {\begin{minipage}[t]{0.33\linewidth}
  \centering
  \includegraphics*[scale = .25]{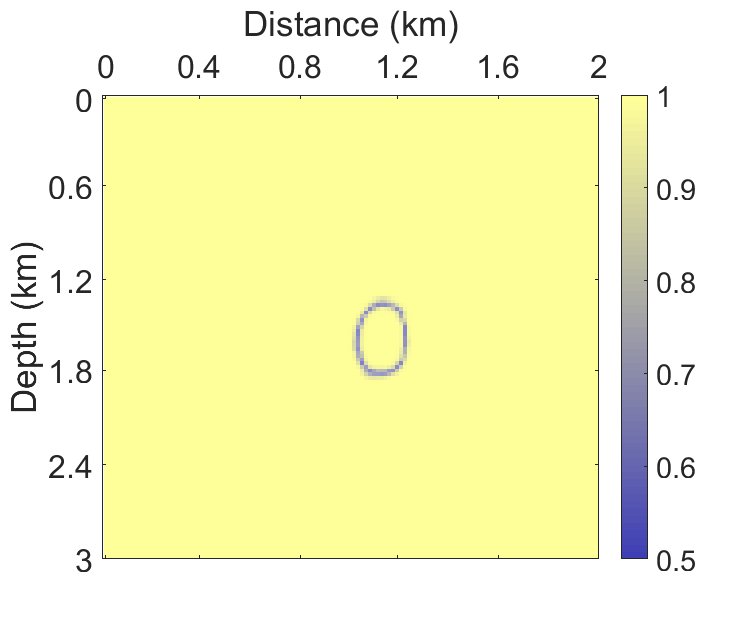}
  \end{minipage}}
\caption{2D pocket results predicted by GeoDUDe-4, with a typical data point chosen for visualization. The pocket is recovered with the networks confidence wavering on the boundary of the pocket.}\label{fig:2dp_results}
\end{figure}

{\bf Results. } The network achieved a training accuracy of 99.95\% and an evaluation accuracy of 99.73\% after 1428 epochs. In Figure \ref{fig:2dp_results} we see the networks are indeed learning geometry. This is particularly interesting given that the network only "sees" images like Figure ~\ref{fig:2dp_disp}(b). These results suggest the network is transforming the data in some way which we hope to explore in future work. 

%%
%\begin{figure}
%  \centering
%  \includegraphics*[scale = .35]{icolor_heat.png}
%\caption{Visualization of network input as image for 2D pocket model. Each color channel (inverse RGB) represents a coordinate of the displacement.}\label{fig:2dp_color}
%\end{figure}
%%
%%
%%

%\subsubsection{Multiple Pocket Domain}
%{\bf Dataset} For this data set change the source to $(.4,1,1.75)$ km.  The interfaces are still fixed at 1km and 2.5km. Three cylinders with center $(x,z)$ and radius $r$ will be randomly generated $x\in [0.85 ,1.65]$km,  $z\in [1.35, 2.15]$km, and $r \in [.05,.3]$. Besides the range of values, there are no other restrictions for the three cylinders, allowing for partial or total overlap, resulting in a wider class of possible configurations and substructure geometries. The radius $r$ is sampled from the probability distribution \textbf{INCLUDE THE DISTRIBUTION}
%
%{\bf Network Design. }
%
%{\bf Results. }
%%%
%%%%%
%%%
%\begin{figure}
  %\subfigure[Actual]
  %{\begin{minipage}[t]{0.33\linewidth}
  %\centering
  %\includegraphics*[scale = .25]{mp_actu.png}
  %\end{minipage}}%%
  %\subfigure[Predicted]
  %{\begin{minipage}[t]{0.33\linewidth}
  %\centering
  %\includegraphics*[scale = .25]{mp_actu.png}
  %\end{minipage}}%%
  %\subfigure[Difference]
  %{\begin{minipage}[t]{0.33\linewidth}
  %\centering
  %\includegraphics*[scale = .25]{mp_actu.png}
  %\end{minipage}}
%\caption{PLACEHOLDER}\label{fig:mp_results}
%\end{figure}
%%%
%%%%%
%%%
\subsection{Effect of Noisy data}\label{sec:noise}
\begin{figure}
  \subfigure[No noise]
  {\begin{minipage}[t]{0.5\linewidth}
  \centering
  \includegraphics*[scale = .35]{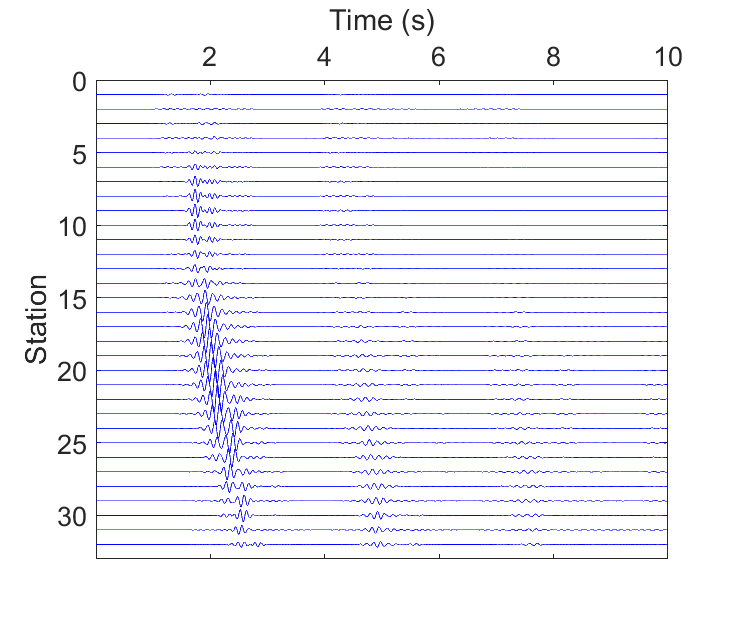}
  \end{minipage}}%%
  %\subfigure[$\sigma$ = 0.002]
  %{\begin{minipage}[t]{0.33\linewidth}
  %\centering
  %\includegraphics*[scale = .25]{2dp_noise_02.png}
  %\end{minipage}}
  \subfigure[Noise strength of 1\% of maximum recorded data]
  {\begin{minipage}[t]{0.5\linewidth}
  \centering
  \includegraphics*[scale = .35]{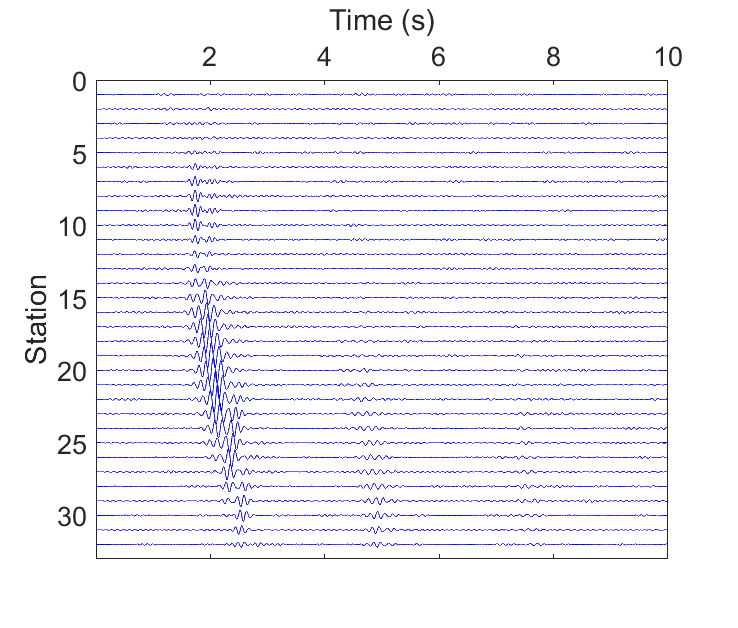}
  \end{minipage}}%%
\caption{Comparison of seismograms with noise and no noise for the 2D pocket model. (a) Seismogram with no noise. (b) Additive Gaussian white noise at $1\%$ of $\max|\bu|$. This shows that $1\%$ of the maximum recorded displacement is enough to mask the reflected data from the pocket.}\label{fig:noise_seis}
\end{figure}

We now consider the 2D pocket example with additive white noise.  Normally, noise is added to the training data set to increase the size of the set and lead to a more robust network. We take an evaluation set of 1000 data points and add {\it i.i.d.} (independent identically distributed) Gaussian noise to each time step of the displacement field data. For an individual data point, the noise strength can be calculated by
\begin{equation}
W_i = \frac{\sigma}{R\max|\bu_r|},
\end{equation}
where $R$ is the reflection coefficient and $\max|\bu_r|$ is the maximum displacement from the reflected wave. The noise strength will be given by $W$, which is the approximate average value of $W_i$ across the data set. The standard deviation $\sigma$ is chosen so that $W$ can be interrupted as a percentage of the reflected wave displacement, e.g., $W = 20$ gives of the a noise strength of 20\% of the average max displacement of the reflected wave. We notice that with noise generated with a strength of 1\% of the maximum of direct recorded displacement, the reflected data from the pocket is the same order of magnitude of the noise, effectively masking it. See Figure~\ref{fig:noise_seis}. \\

{\bf Network Design:} To compare results, we use the same model as in the previous Section~\ref{2dp:section} and train a network with the same parameters, with a noise strength of $W = 20$. \\

\begin{figure}
  \subfigure[ground truth]
  {\begin{minipage}[t]{0.33\linewidth}
  \centering
  \includegraphics*[scale = .25]{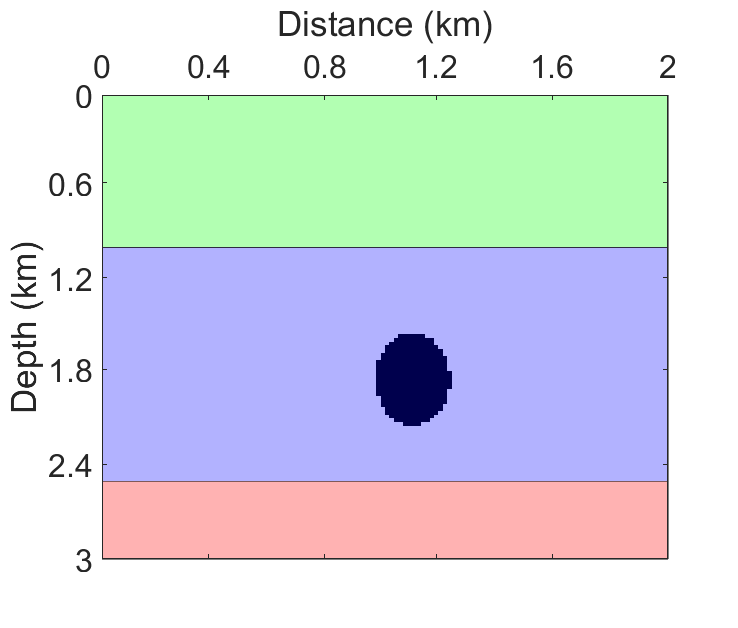}
  \end{minipage}}%%
  %\subfigure[OUI=0.1403, PDC=0.8910]
  \subfigure[IOU=0.1403]
  {\begin{minipage}[t]{0.33\linewidth}
  \centering
  \includegraphics*[scale = .25]{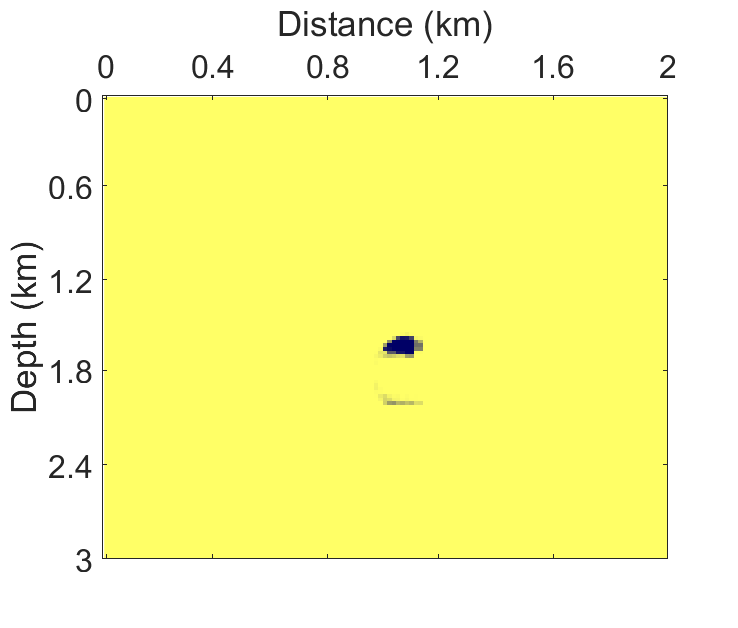}
  \end{minipage}}%%
  %\subfigure[IOU=0.2052, PDC=0.6261]
  \subfigure[IOU=0.2052]
  {\begin{minipage}[t]{0.33\linewidth}
  \centering
  \includegraphics*[scale = .25]{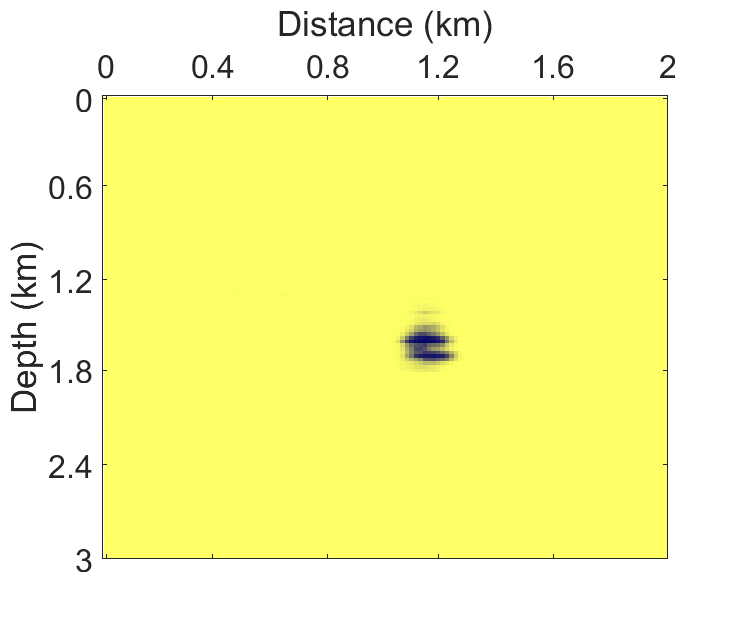}
  \end{minipage}} \\
\caption{Visualization of IOUs by GeoDUDe-4 for the 2D pocket model. Results taken from network trained with noise. Data is augmented with noise with a noise strength of 50\%. (a) ground truth for comparision. (b) IOU=0.1403. (c) IOU= 0.2052. For each displayed results, the networks are able to detect the location of the pocket. With additional noise the network is unable to resolve the geometry.}\label{fig:iou_pdc}
\end{figure}

\begin{figure}
  \subfigure[No noise]
  {\begin{minipage}[t]{0.33\linewidth}
  \centering
  \includegraphics*[scale = .25]{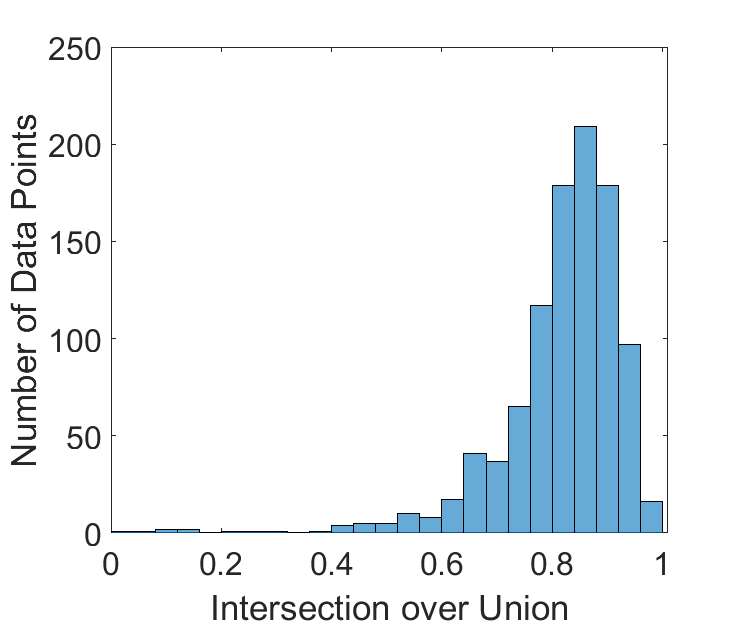}
  \end{minipage}}%%
  \subfigure[W=10]
  {\begin{minipage}[t]{0.33\linewidth}
  \centering
  \includegraphics*[scale = .25]{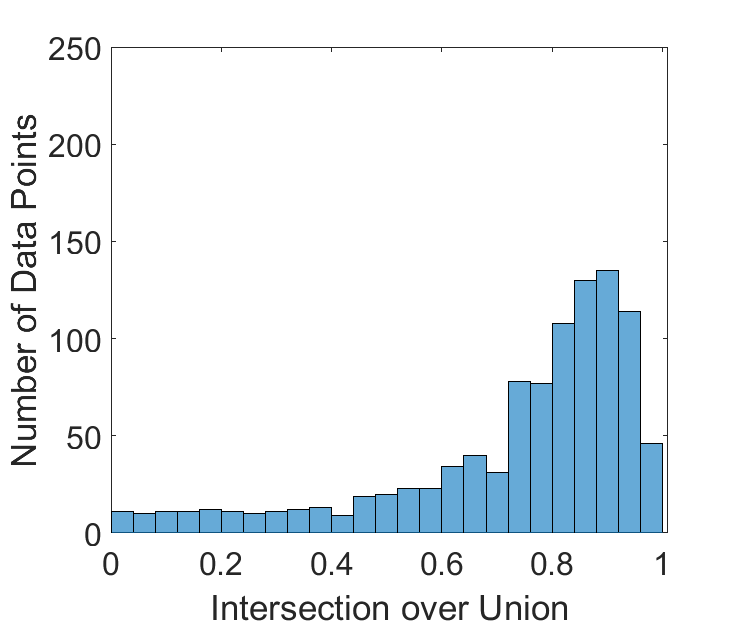}
  \end{minipage}}%%
  \subfigure[W=50]
  {\begin{minipage}[t]{0.33\linewidth}
  \centering
  \includegraphics*[scale = .25]{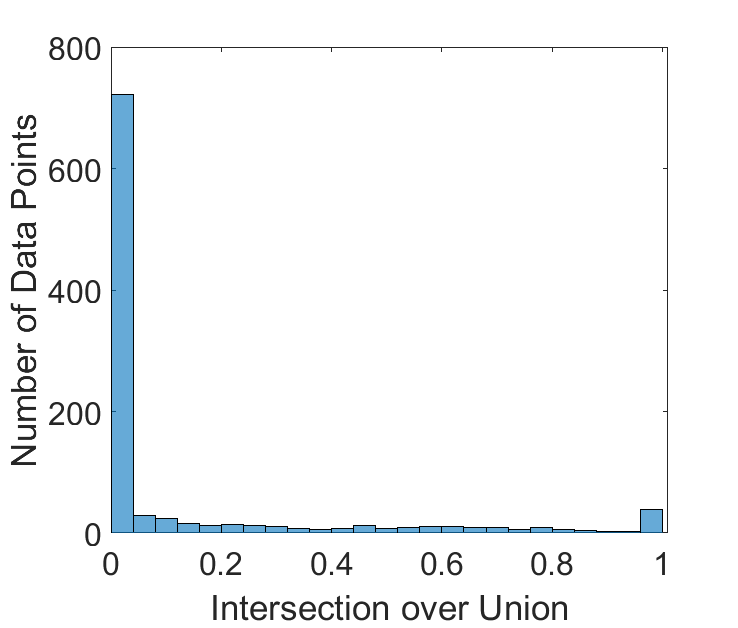}
  \end{minipage}} \\
	%  \subfigure[No noise]
  %{\begin{minipage}[t]{0.33\linewidth}
  %\centering
  %\includegraphics*[scale = .25]{2dp_00_pdc.png}
  %\end{minipage}}%%
  %\subfigure[10\%]
  %{\begin{minipage}[t]{0.33\linewidth}
  %\centering
  %\includegraphics*[scale = .25]{2dp_01_pdc.png}
  %\end{minipage}}%%
  %\subfigure[50\%]
  %{\begin{minipage}[t]{0.33\linewidth}
  %\centering
  %\includegraphics*[scale = .25]{2dp_05_pdc.png}
  %\end{minipage}} \\
\caption{Network with trained without noise, 1000 data points are plotted in each Histogram. Subfigures (a), (b), (c) show the IOU metric
%, subfigures (d), (e), (f) show the PDC metric with ,
with no noise, 10\% noise strength, and 50\% noise strength respectively.}\label{fig:2dp_nonoise_hist}
\end{figure}

{\bf Results. } %We report 2 metrics, which can be seen in figs.~\ref{fig:2dp_nonoise_hist},\ref{fig:2dp_noise_hist}.  
A GeoDUDe-4 was trained for 2000 epochs with additional noise for a final evaluation accuracy of 99.731\% evaluation accuracy. However, evaluation accuracy can be a misleading metric for network performance in pocket detection since assigning the high velocity class to every pixel could get an accuracy up to 80\% on some samples. Instead intersection over union (IOU) is used (see \cite{rahman2016} for a more detailed explanation). Figures \ref{fig:iou_pdc} and \ref{fig:2dp_nonoise_hist} show the histograms the IOU scores of networks trained with and without white noise evaluated on the evaluation data with no additional noise, additional noise strength $W= 10$, and additional noise strength $W = 50$ respectively. While both networks display good IOU scores on the unperturbed data and when the data is only perturbed with noise strength $W = 10$, the benefits of additional noise in training become clear when the noise strength is increased to $W = 50$: the IOU scores of the network trained without noise on noisy data plummets, effectively misclassifying almost every pocket, while the IOU score of the network trained with noise decreases, but maintains many correct classifications. The average IOU scores are summarized in Table \ref{tab:iou_comparison}. Evaluating on higher noise strength collapses the network's output to no pocket detected.

%\begin{table}
	%\begin{center}
		%\label{tab:iou_comparison}
		%\begin{tabular}{c|c|c|c}
			 %& \textbf{Unperturbed} & \textbf{Perturbed by $W = 10$} & \textbf{Perturbed by $W = 50$} \\
			%\hline
			%\textbf{Trained without Noise} & 0.9453 & 0.8650  & 0.1383  \\
			%\hline
			%\textbf{Trained with Noise} & 0.9155 & 0.8776 & 0.6025 \\
		%\end{tabular}
		%\caption{IOU Scores for GeoDUDe-4 trained with and without noise.}
	%\end{center}
%\end{table}

\begin{table}
	\begin{center}
		\begin{tabular}{c|c|c|c}
			 & \textbf{Unperturbed} & \textbf{Perturbed by $W = 10$} & \textbf{Perturbed by $W = 50$} \\
			\hline
			\textbf{Trained without Noise} & 0.8163 & 0.7335  & 0.1308  \\
			\hline
			\textbf{Trained with Noise} & 0.8706 & 0.7576 & 0.5249 \\
		\end{tabular}
		\caption{IOU Scores for GeoDUDe-4 trained with and without noise for the 2D pocket model.}\label{tab:iou_comparison}
	\end{center}
\end{table}

%The other metric is what we call pocket detection confidence (PDC), it measures of confidence per pixel for the low velocity region. A score on this metric near one means the network detected a pocket with good confidence, a score near zero means no pocket was detected.  With these 2 metrics we can quantify how the networks are performing; e.g. A high IOU score and high PDC score means the network detected a pocket in the correct place. A low IOU score and high PDC score means the network detected a pocket incorrectly, meaning the noise was strong enough for the network to shift the pocket. A high IOU and low PDC score, is interpreted as the network can detect the pocket but it is not confident it is there.  A low IOU score and low PDC score, means the noise is to strong and the network could not find the pocket, so it returns no pocket. Figure~\ref{fig:iou_pdc} gives a visualization of what these two metrics measure.

%\begin{figure}
  %\subfigure[No noise]
  %{\begin{minipage}[t]{0.33\linewidth}
  %\centering
  %\includegraphics*[scale = .25]{2dp_0_pred_n.png}
  %\end{minipage}}%%
  %\subfigure[0.1\%]
  %{\begin{minipage}[t]{0.33\linewidth}
  %\centering
  %\includegraphics*[scale = .25]{2dp_01_pred_n.png}
  %\end{minipage}}%%
  %\subfigure[0.3\%]
  %{\begin{minipage}[t]{0.33\linewidth}
  %\centering
  %\includegraphics*[scale = .25]{2dp_03_pred_n.png}
  %\end{minipage}} \\
	  %\subfigure[0.5\%]
  %{\begin{minipage}[t]{0.33\linewidth}
  %\centering
  %\includegraphics*[scale = .25]{2dp_05_pred_n.png}
  %\end{minipage}}%%
  %\subfigure[1\%]
  %{\begin{minipage}[t]{0.33\linewidth}
  %\centering
  %\includegraphics*[scale = .25]{2dp_10_pred_n.png}
  %\end{minipage}}%%
  %\subfigure[2\%]
  %{\begin{minipage}[t]{0.33\linewidth}
  %\centering
  %\includegraphics*[scale = .25]{2dp_20_pred_n.png}
  %\end{minipage}} \\
%\caption{Network with trained noise: Confidence map: White noise is with a strength percentage of maximum value of the recorded wavefield}\label{fig:2dp_noise_heat}
%\end{figure}

\begin{figure}
  \subfigure[No noise]
  {\begin{minipage}[t]{0.33\linewidth}
  \centering
  \includegraphics*[scale = .25]{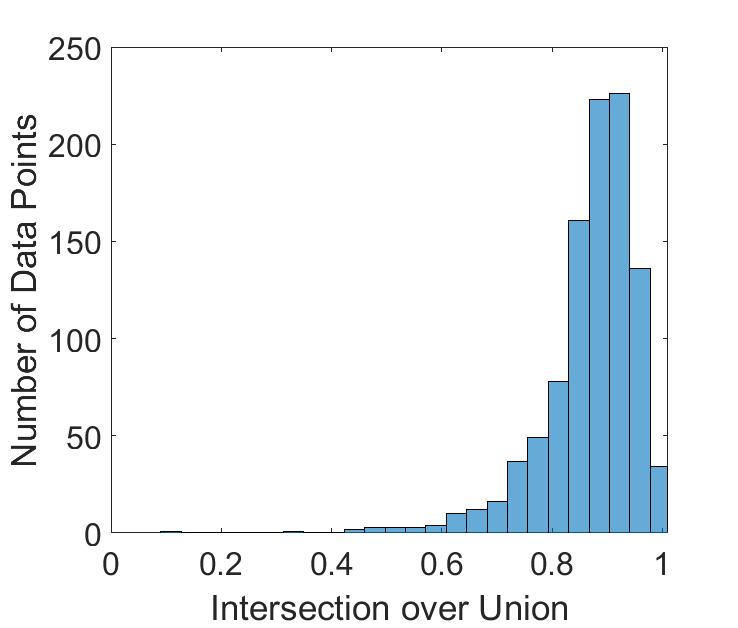}
  \end{minipage}}%%
  \subfigure[W=10]
  {\begin{minipage}[t]{0.33\linewidth}
  \centering
  \includegraphics*[scale = .25]{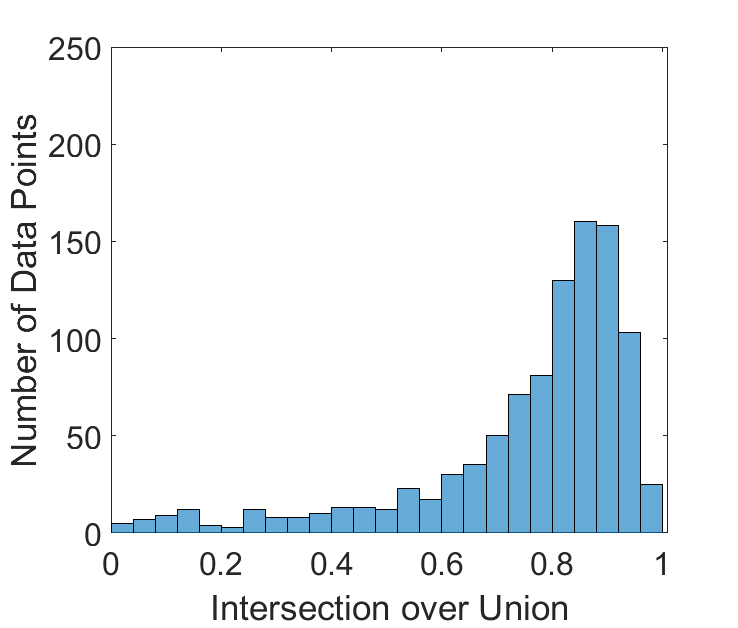}
  \end{minipage}}%%
  \subfigure[W=50]
  {\begin{minipage}[t]{0.33\linewidth}
  \centering
  \includegraphics*[scale = .25]{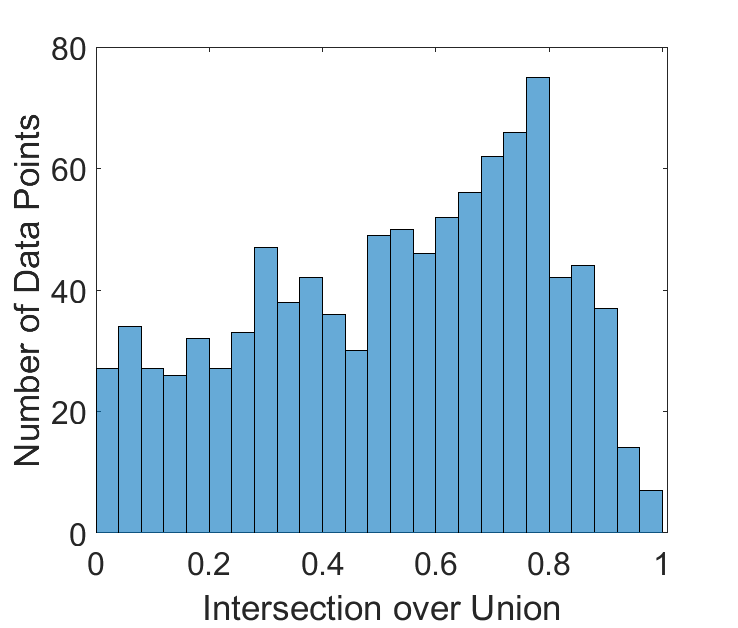}
  \end{minipage}} \\
%	  \subfigure[No noise]
  %{\begin{minipage}[t]{0.33\linewidth}
  %\centering
  %\includegraphics*[scale = .25]{2dp_00_pdc_n.png}
  %\end{minipage}}%%
  %\subfigure[10\%]
  %{\begin{minipage}[t]{0.33\linewidth}
  %\centering
  %\includegraphics*[scale = .25]{2dp_01_pdc_n.png}
  %\end{minipage}}%%
  %\subfigure[50\%]
  %{\begin{minipage}[t]{0.33\linewidth}
  %\centering
  %\includegraphics*[scale = .25]{2dp_05_pdc_n.png}
  %\end{minipage}} \\
\caption{Performance of the Network GeoDUDe-4 trained with noise strength at 20\% of the average max displacement of the reflected wave for the 2D pocket model. 1000 data points are plotted in each Histogram. Subfigures (a), (b), (c) show the IOU metric, 
%subfigures (d), (e), (f) show the PDC metric with no added noise,
with no noise, 20\% noise strength, and 50\% noise strength, respectively.
}\label{fig:2dp_noise_hist}
\end{figure}

\subsection{Structured noise}
{We now consider the 2D pocket example with additive structured noise. From our noiseless evaluation data set, we add structured low frequency noise to each receiver.  Numbering each receiver, $S_j = (j2/31,1,0)$, for $j = 0,\ldots, 31$ we add noise to receiver $S_j$ as
\begin{align}\label{eq:struct_noise_freq}
\frac{a}{2}\cos(2t)\frac{4}{3(32 - j)},
\end{align}
where the amplitude $a$ is modulus of the maximum displacement of the wavefield reflected from the interface. We add this noise to each component of the wavefield.  We remark that at receiver at $x = 2$, the noise is the strongest at with an amplitude of $2a/3$.  That is, the noise strength is two thirds of the height of the modulus of the maximum displacement wavefield reflected from the interface.  This is a stronger noise than the $W = 50$ case of additive white noise. The structured noise decreases to an amplitude of $a/48$ at the receiver located at $x = 0$
}

\begin{figure} 
  \subfigure[Actual]
  {\begin{minipage}[t]{0.33\linewidth}
  \centering
  \includegraphics*[scale = .25]{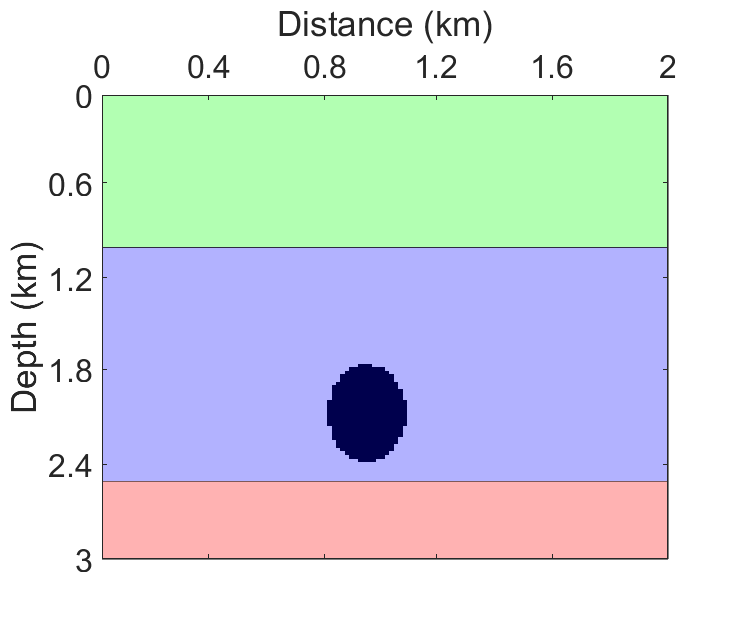}
  \end{minipage}}%%
  \subfigure[Predicted,  IOU=0.7585]
  {\begin{minipage}[t]{0.33\linewidth}
  \centering
  \includegraphics*[scale = .25]{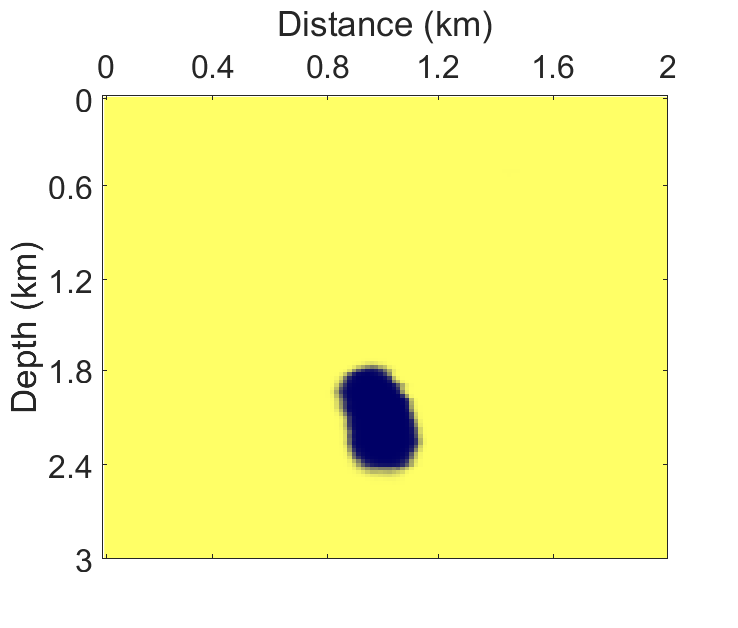}
  \end{minipage}}
  \subfigure[Confidence map]
  {\begin{minipage}[t]{0.33\linewidth}
  \centering
  \includegraphics*[scale = .25]{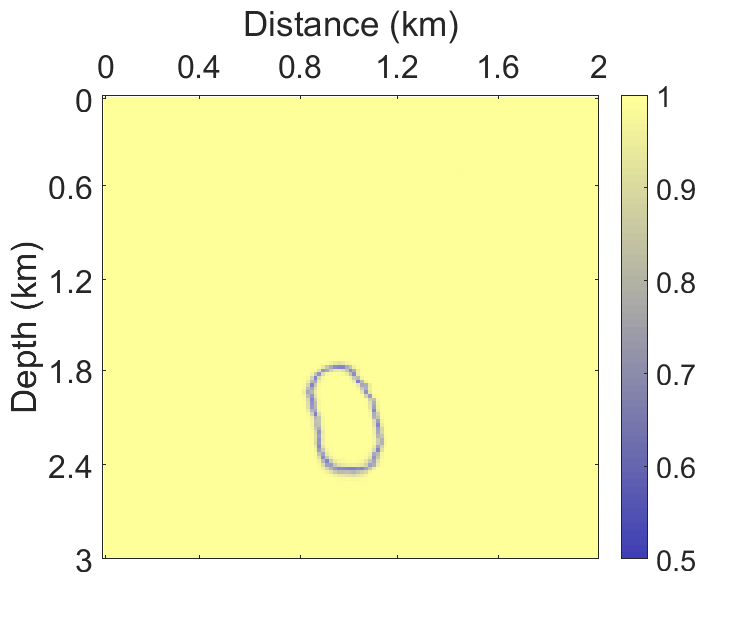}
  \end{minipage}}%%
\caption{A visualization of a result from the structured noise~\eqref{eq:struct_noise_freq} added to the noiseless evaluation dataset. The network used is GeoDUDe-4 trained with noise from section 4.5. (a) Ground truth. (b) Prediction, with an IOU value of 0.7585. (c) Confidence map. }\label{fig:struct_noise}
\end{figure}

\begin{figure}
\centering
  %\subfigure[]
  {\begin{minipage}[t]{0.5\linewidth}
  \includegraphics*[scale = .4]{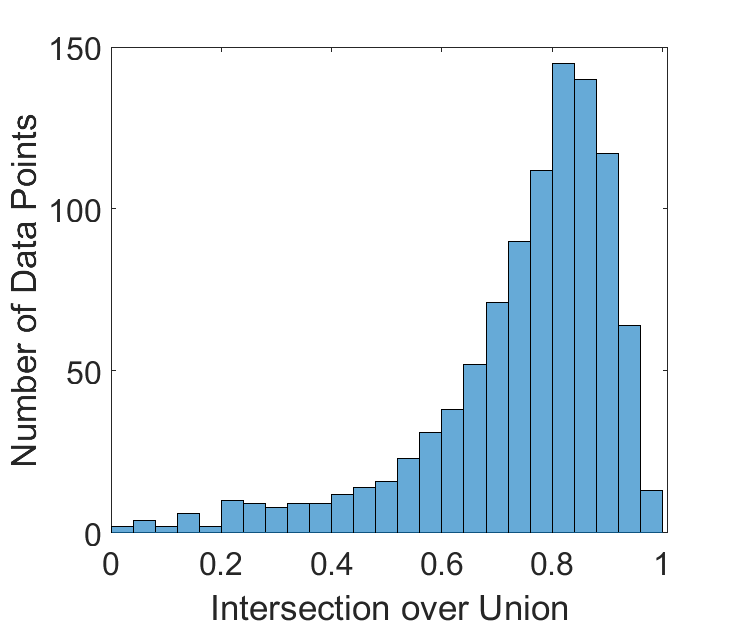}
  \end{minipage}}%%
\caption{The IOU metric for structured noise as the input into GeoDUDe-4 for the 2D pocket model trained with noise from section 4.5. The average IOU is $.7443$.}\label{fig:struct_noise_hist}
\end{figure}

{
The network is able to detect the pocket with good success.  The network has full confidence, as can be seen from Fig.~\ref{fig:struct_noise}c; however, the boundary for the prediction is perturbed. This gives a slight false result which is reflected in the IOU metric, see Fig.~\ref{fig:struct_noise_hist}. With the structured noise, the prediction has a lower average IOU value of $.7443$ compared to $0.8706$ with no noise.

}

\section{Conclusions and future work}\label{sec:conclusion}
The use of the FGA to generate large amounts of seismic data provides a quick way to generate labeled synthetic data for statistical learning of the inverse tomography problem. Casting the inverse problem as a segmentation problem resulted in high evaluation accuracy networks for piecewise constant two-layer models on both FGA and SEM datasets. The
%Encode/Decode fully convolutional and 
UNet architectures with dense blocks displayed superior accuracy compared to simpler network architectures, however, deeper networks did not necessarily outperform their shorter counterparts. On the two layer benchmark problem the networks exhibited good invariance of prediction in regard to which numerical method was used to generate the dataset, likely because the FGA and SEM exhibit the same traveltime information. Having a network independent of numerical method is important, and the FGA can help to train such a network as it generates synthetic seismic data that carries the correct traveltime information of the real-world data. Further, analogous meta-architectures also exhibit high evaluation and IOU accuracy for pocket detection in noisy data.

The success of the networks on the substructure geometries in the paper act as a stepping stone to tackle more complicated and realistic geological models. By developing the API GeoSeg, available at \href{https://github.com/KyleMylonakis/GeoSeg}{https://github.com/KyleMylonakis/GeoSeg}, it is easy to implement neural networks designed for the reported example models and more general segmentation problems of seismogram data than those discussed in this paper. Together with the FGA, the task of training a deep neural network on sufficiently large amounts of seismogram data becomes a computationally affordable task. Immediate future directions to be explored are multi-pocket models, multi-nonlinear interface models with and without pockets present. Long term goal is to develop a neural network model to tackle fully 3D substructure geometries and develop a neural network trained on synthetic seismic data capable of making inferences from real seismic data.
\section*{Acknowledgement}
We acknowledge support from the Center for Scientific Computing from the CNSI, MRL: an NSF MRSEC (DMR-1720256) and NSF CNS-1725797. The work was partially supported by the NSF grant DMS-1818592. XY also thanks Professors Haizhao Yang and Kui Ren for useful discussions.
%---APPENDIX----------------------------------------------------------------------------------
%\appendix
%\label{sec:append}
%\nocite{*}
%\clearpage
%\newpage
%---REFERENCES--------------------------------------------------------------------------------
\bibliographystyle{abbrv}
\bibliography{bib_dl_fga,fga_ref,SEG}

\begin{thebibliography}{10}

\bibitem{Aki1976}
K.~Aki and W.~Lee.
\newblock {Determination of the three-dimensional velocity anomalies under a
  seismic array using first P arrival times from local earthquakes 1. A
  homogeneous intial model}.
\newblock {\em J. Geophys. Res.}, 81:4381--4399, 1976.

\bibitem{doi:10.1190/tle36030208.1}
M.~Araya-Polo, T.~Dahlke, C.~Frogner, C.~Zhang, T.~Poggio, and D.~Hohl.
\newblock Automated fault detection without seismic processing.
\newblock {\em The Leading Edge}, 36(3):208--214, 2017.

\bibitem{araya2018deep}
M.~Araya-Polo, J.~Jennings, A.~Adler, and T.~Dahlke.
\newblock Deep-learning tomography.
\newblock {\em The Leading Edge}, 37(1):58--66, 2018.

\bibitem{rahman2016}
M.~Atiqur~Rahman and Y.~Wang.
\newblock Optimizing intersection-over-union in deep neural networks for image
  segmentation.
\newblock volume 10072, pages 234--244, 12 2016.

\bibitem{chai2017frozen}
L.~Chai, P.~Tong, and X.~Yang.
\newblock Frozen {G}aussian approximation for 3-{D} seismic wave propagation.
\newblock {\em Geophysical Journal International}, 208(1):59--74, 2017.

\bibitem{dozat2016incorporating}
T.~Dozat.
\newblock Incorporating nesterov momentum into adam.
\newblock 2016.

\bibitem{dziewonski1981preliminary}
A.~M. Dziewonski and D.~L. Anderson.
\newblock Preliminary reference earth model.
\newblock {\em Physics of the earth and planetary interiors}, 25(4):297--356,
  1981.

\bibitem{Glorot2011DeepSR}
X.~Glorot, A.~Bordes, and Y.~Bengio.
\newblock Deep sparse rectifier neural networks.
\newblock In {\em AISTATS}, 2011.

\bibitem{hateley:2018}
J.~C. Hateley, X.~Yang, L.~Chai, and P.~Tong.
\newblock {Frozen Gaussian approximation for 3-D elastic wave equation and
  seismic tomography}.
\newblock {\em Geophysical Journal International}, 216(2):1394--1412, 11 2019.

\bibitem{He_Zhang_Ren_Sun_ResNets}
K.~He, X.~Zhang, S.~Ren, and J.~Sun.
\newblock Deep residual learning for image recognition.
\newblock {\em CoRR}, abs/1512.03385, 2015.

\bibitem{he2016deep}
K.~He, X.~Zhang, S.~Ren, and J.~Sun.
\newblock Deep residual learning for image recognition.
\newblock In {\em Proceedings of the IEEE conference on computer vision and
  pattern recognition}, pages 770--778, 2016.

\bibitem{Huang_Liu_Weinberger_DN}
G.~Huang, Z.~Liu, and K.~Q. Weinberger.
\newblock Densely connected convolutional networks.
\newblock {\em CoRR}, abs/1608.06993, 2016.

\bibitem{Ioffe_Szegedy_BN}
S.~Ioffe and C.~Szegedy.
\newblock Batch normalization: Accelerating deep network training by reducing
  internal covariate shift.
\newblock {\em CoRR}, abs/1502.03167, 2015.

\bibitem{DBLP:journals/nature/LeCunBH15}
Y.~LeCun, Y.~Bengio, and G.~E. Hinton.
\newblock Deep learning.
\newblock {\em Nature}, 521(7553):436--444, 2015.

\bibitem{LuYa:MMS}
J.~Lu and X.~Yang.
\newblock Frozen {G}aussian approximation for general linear strictly
  hyperbolic systems: {F}ormulation and {E}ulerian methods.
\newblock {\em Multiscale Model. Simul.}, 10:451--472, 2012.

\bibitem{AO_Fawzi_Frossard_UniAdPert}
S.~Moosavi{-}Dezfooli, A.~Fawzi, O.~Fawzi, and P.~Frossard.
\newblock Universal adversarial perturbations.
\newblock {\em CoRR}, abs/1610.08401, 2016.

\bibitem{nakamichi2003source}
H.~Nakamichi, H.~Hamaguchi, S.~Tanaka, S.~Ueki, T.~Nishimura, and A.~Hasegawa.
\newblock Source mechanisms of deep and intermediate-depth low-frequency
  earthquakes beneath iwate volcano, northeastern japan.
\newblock {\em Geophysical Journal International}, 154(3):811--828, 2003.

\bibitem{Perole1700578}
T.~Perol, M.~Gharbi, and M.~Denolle.
\newblock Convolutional neural network for earthquake detection and location.
\newblock {\em Science Advances}, 4(2), 2018.

\bibitem{Rawlinson2010}
N.~Rawlinson, S.~Pozgay, and S.~Fishwick.
\newblock {Seismic tomography: A window into deep Earth}.
\newblock {\em Phys. Earth Planet. Inter.}, 178(3-4):101--135, 2010.

\bibitem{Romanowicz1991}
B.~Romanowicz.
\newblock {Seismic tomography of the Earth's mantle}.
\newblock {\em Annu. Rev. Earth Planet. Sci.}, 19:77--99, 1991.

\bibitem{Ronn_Fischer_UNet}
O.~Ronneberger, P.~Fischer, and T.~Brox.
\newblock U-net: Convolutional networks for biomedical image segmentation.
\newblock {\em CoRR}, abs/1505.04597, 2015.

\bibitem{Shelhamer2015FullyCN}
E.~Shelhamer, J.~Long, and T.~Darrell.
\newblock Fully convolutional networks for semantic segmentation.
\newblock {\em 2015 IEEE Conference on Computer Vision and Pattern Recognition
  (CVPR)}, pages 3431--3440, 2015.

\bibitem{SpringenbergDBR14}
J.~T. Springenberg, A.~Dosovitskiy, T.~Brox, and M.~A. Riedmiller.
\newblock Striving for simplicity: The all convolutional net.
\newblock {\em CoRR}, abs/1412.6806, 2014.

\bibitem{Szegedy_Zaremba_Sutskever_intriguingNN}
C.~Szegedy, W.~Zaremba, I.~Sutskever, J.~Bruna, D.~Erhan, I.~J. Goodfellow, and
  R.~Fergus.
\newblock Intriguing properties of neural networks.
\newblock {\em CoRR}, abs/1312.6199, 2013.

\bibitem{WeYa:12}
D.~Wei and X.~Yang.
\newblock Eulerian gaussian beam method for high frequency wave propagation in
  heterogeneous media with discontinuities in one direction.
\newblock {\em Commun. Math. Sci.}, 10:1287--1299, 2012.

\bibitem{8424545}
Y.~Wu, Y.~Lin, Z.~Zhou, D.~C. Bolton, J.~Liu, and P.~Johnson.
\newblock Deepdetect: A cascaded region-based densely connected network for
  seismic event detection.
\newblock {\em IEEE Transactions on Geoscience and Remote Sensing}, pages
  1--14, 2018.

\bibitem{Yi:01}
O.~Yilmaz.
\newblock {\em Seismic Data Analysis: Processing, Inversion, and Interpretation
  of Seismic Data}.
\newblock Society of Exploration Geophysicists, 2001.

\bibitem{Zhang_Liu_RUNet}
Z.~Zhang, Q.~Liu, and Y.~Wang.
\newblock Road extraction by deep residual u-net.
\newblock {\em CoRR}, abs/1711.10684, 2017.

\bibitem{Zhao2012a}
D.~Zhao.
\newblock {Tomography and dynamics of Western-Pacific subduction zones}.
\newblock {\em Monogr. Environ. Earth Planets}, 1:1--70, 2012.

\bibitem{2018arXiv180303211Z}
W.~{Zhu} and G.~C. {Beroza}.
\newblock {PhaseNet: A Deep-Neural-Network-Based Seismic Arrival Time Picking
  Method}.
\newblock {\em ArXiv e-prints}, Mar. 2018.

\end{thebibliography}

%\newpage
%\resetfigno

\end{document}